# A Mechanism-Based Planning Framework for Equitable and Merit-Preserving University Admissions

**Jung-Ah Lee**

**ORCID:** 0009-0003-1278-7912

Independent Researcher, Seoul, Republic of Korea

ljak96@seoultech.ac.kr; ava.jahlee@gmail.com

## Abstract

Admissions systems in many countries struggle to balance merit-based selection with equity objectives. Most existing approaches—categorical quotas, fragmented equity tracks, and opaque adjustments—lack transparent decision rules and operational coherence. This paper introduces the **Adaptive Merit Framework (AMF)**, a mechanism-based architecture that combines an individual-level SES correction rule with a structured decision pipeline. AMF operates under a non-displacement constraint: regular admissions remain determined entirely by raw merit scores, and only applicants whose corrected performance exceeds the same threshold quality as conditional admits. The framework is operationalized through a five-stage decision spine—input definition, indicator aggregation, equity calibration via a single parameter $\alpha$, batch execution, and irreversible closure—eliminating institutional discretion throughout. An empirical application using PISA 2022 Korea data (N = 6,377) shows that AMF identifies 4-9 additional candidates exclusively from the bottom half of the SES distribution, all above the merit threshold, expanding admissions by fewer than 0.15% of the cohort. The results demonstrate that rule-based correction can recover suppressed high-merit individuals without displacing standard admits, providing a transparent and scalable alternative to discretionary equity interventions.

## 1. Introduction

### 1.1. From Inequality of Opportunity to Talent Recognition

Social mobility—the capacity for individuals to improve their socioeconomic position beyond that of their parents—is a foundational element of democratic legitimacy. Yet recent evidence indicates that mobility in Korea has become increasingly rigid. According to KRIVET [1], intergenerational mobility has declined steadily over the past fifteen years. This trend aligns with Korea's high intergenerational income elasticity (IGE = 0.60) [2], which reflects the strong persistence of inequality across generations.

What makes this rigidity particularly striking is that it emerges despite Korea's well-documented educational strength. Korea ranks among the top performers in the OECD's **PISA 2022 Creative Thinking** assessment, and socioeconomic status (SES) explains only 6.4% of the variance in scores-well below the OECD average of 11.6%. Analysis of the raw PISA 2022 dataset further reveals that, despite substantial average performance gaps, a nontrivial minority of low-SES students achieve high academic outcomes, representing a population of "**hidden excellence**"—individuals who perform strongly despite structural disadvantage.

However, this demonstrated **academic development does not translate into upward mobility**. KRIVET [1] reports a 22% decline in upward mobility (from 13.4% to 10.5%), with mathematics-based mobility now among the lowest in comparable economies. In effect, Korea has built an education system that successfully develops cognitive potential but does not consistently convert that potential into opportunity. **A strength in talent development becomes a weakness in talent recognition.**

Although social mobility is influenced by broader economic structures, education remains the most immediate institutional lever for improving opportunity. The challenge today is therefore not whether fairness interventions are necessary, but *how they can be designed to correct structural bias while preserving meritocratic standards*.

### 1.2. The Fairness Paradox in College Admissions

College admissions systems around the world reveal a persistent tension between **meritocracy** and **equity**. Standardized test scores are often viewed as objective indicators of merit, yet they systematically reflect socioeconomic advantage. Conversely, policies designed to correct these disparities, such as affirmative action, face accusations of reverse discrimination and a lack of transparency, undermining public trust.

This tension is not merely political but conceptual. As Sandel argues, judgments about fairness in admissions cannot be separated from the purpose of the institution itself [3]. If that purpose includes recognizing latent potential—the ability to achieve under conditions of equal opportunity—then socioeconomic context becomes a necessary input, not an optional adjustment. And if so, the selection rule itself must be explicitly designed to reflect that purpose, rather than left to institutional discretion. Recent study shows that threshold-based policies can improve short-run efficiency while worsening equity [4]. This motivates equity-enhancing mechanisms that explicitly avoid displacement.

Recent events in Korea illustrate how fairness interventions can lose legitimacy when their design lacks transparency. In particular, the **regional-balance** track used by Seoul National University and other institutions has faced criticism despite its stated goal of promoting geographic diversity. Between 2020 and 2024, the proportion of admits from Seoul metropolitan area high schools increased from 51.6% (2020) to 61.5% (2024) [5], raising concerns about proxy-based targeting and unclear selection criteria. Notably, regional-balance admits achieved higher average graduation GPAs than regular admits (3.67 vs. 3.61), suggesting that **the issue lies not in the idea of fairness interventions but in their design**.

Comparable patterns appear internationally. In the United States, decades of race-based affirmative action culminated in the Supreme Court's 2023 decision [6], which ruled such practices unconstitutional, emphasizing that coarse group-based proxies can misclassify disadvantage and undermine legitimacy. Similarly, the College Board's short-lived **Adversity Score** (2019) drew criticism for relying on neighborhood- and school-level indicators and for its opaque scoring methodology, highlighting the risks of proxy dependence and non-transparent calibration in fairness-oriented tools.

Together, these cases reveal three recurring structural challenges in the design of existing fairness interventions:

- **Proxy dependence**
- **Zero-sum structure**
- **Opacity**

If admissions explicitly adopt a potential-oriented purpose, the central question becomes *how to construct mechanisms that recognize potential without undermining merit.*

### 1.3. Beyond Zero-Sum: An Adaptive Merit Framework (AMF)

This paper introduces the Adaptive Merit Framework (AMF), a policy-engineered admissions mechanism designed to recognize latent potential while preserving merit-based standards. Rather than treating demonstrated ability and structural disadvantage as competing priorities, AMF integrates both through a transparent correction model anchored to a dynamic merit threshold.

AMF is built on three structural commitments that distinguish it from existing fairness mechanisms:

First, a **non-displacement structure**: regular admissions remain determined exclusively by raw merit. SES-based corrections create only additional seats beyond the existing quota, ensuring that no regular admit is displaced. Fairness becomes additive rather than substitutive.

Second, a **dynamic threshold**: the merit cutoff is defined by the raw score of the last regular admit, adapting automatically to each applicant pool rather than relying on static or arbitrary standards.

Third, **direct SES measurement**: unlike group-based models that rely on race or region as proxies, AMF operates on individual-level socioeconomic signals drawn from verifiable administrative sources.

These three components are formally specified in **Section 3**, where the correction rule, threshold construction, and calibration procedure are defined.

### 1.4. From Framework to Decision Architecture

While **Section 3 to 4** formalize the correction mechanism and demonstrate its empirical behavior, a key question remains: how should AMF be embedded within real admissions systems constrained by budgets, capacities, and administrative rules? **Section 5** addresses this by introducing a decision spine—an invariant sequence of stages that separates institutional merit construction, policy-calibrated equity adjustment, and execution under capacity constraints. This architecture confines all normative discretion to a single calibration parameter ($\alpha$), while all other elements operate mechanically under transparent rules (see **Fig. 1** for a graphical summary and **Fig. 4** for the detailed architecture).

### 1.5. Contributions

This study contributes to the literature on equity-enhancing admissions mechanisms in three ways. First, it reframes fairness in college admissions as a procedural design problem centered on transparency and legitimacy, rather than as a moral or distributive judgment. Second, it introduces a non-displacement, threshold-anchored selection mechanism that expands opportunity without altering existing merit standards. Third, it specifies a governance-ready decision architecture—the decision spine—that constrains discretionary intervention while maintaining institutional autonomy and administrative feasibility.

### 1.6. Paper Organization

**Fig. 1** provides a graphical summary of the paper structure, tracing the logic from problem identification through mechanism design, empirical validation, and decision architecture. The remainder of this paper follows this progression; supplementary derivations, extended results, and implementation modules are provided in **Appendices A-F** (online supplement).

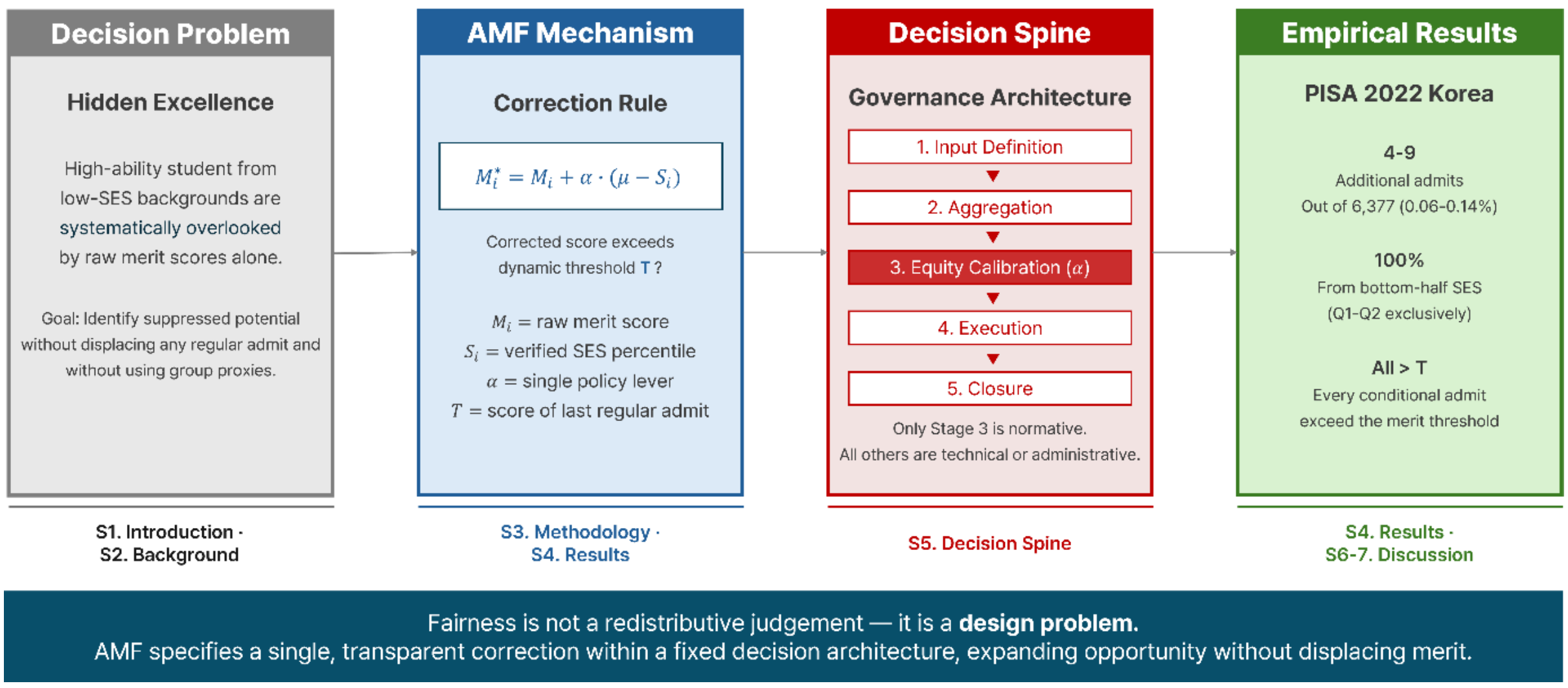

**Figure 1. Graphical summary of the Adaptive Merit Framework (AMF).** The figure traces the logic of the paper from left to right: the decision problem that motivates AMF (left), the correction mechanism (center-left), the decision architecture that governs its implementation (center-right), and the empirical results (right).

## 2. Theoretical Background: Design Constraints in Existing Approaches

This section reviews how prior literatures have operationalized fairness in admissions and related selection contexts, with attention to the design constraints that arise when normative principles are translated into executable decision systems. Rather than reiterating the social motivations for equity discussed in **Section 1**, the review focuses on how prior approaches have specified—or failed to specify—decision rules under legal, and strategic constraints. This perspective clarifies why mechanism-level intervention is required.

### 2.1. Group-Based Interventions and the Limits of Proxy-Based Fairness

Early affirmative action policies sought to address structural inequality through group-based preferences, using race, region, or other categorical attributes as proxies for disadvantage [7]. Such approaches were historically effective in expanding access but increasingly encountered legal and political resistance as admissions systems moved toward individualized evaluation standards [6,8,9].

A core limitation of proxy-based fairness lies in its indirectness. Group membership provides only a coarse signal of individual disadvantage, leading to both inclusion errors and exclusion errors that weaken policy legitimacy. As legal scrutiny intensified, the reliance on categorical proxies became institutionally fragile, constraining the scope of group-based correction even where structural inequality persisted.

Comparative evidence from non-U.S. contexts similarly illustrates the instability of proxy-based admissions. Regional balance and special-track policies have repeatedly faced challenges in accurately targeting disadvantaged individuals, particularly as demographic composition and applicant strategies evolve [5]. Together, these developments underscore a persistent design constraint: fairness mechanisms that rely on indirect proxies struggle to deliver durable, precise correction under contemporary legal and institutional conditions.

### 2.2. Algorithmic Fairness and the Absence of Ex Ante Decision Rules

In parallel, the algorithmic fairness literature has formalized equity concerns through statistical criteria applied to predictive systems, including **demographic parity, equalized odds, individual fairness** [10,11]. These contributions clarify important trade-offs among fairness definitions but typically operate as *ex-post* constraints on model outputs rather than as specifications of decision procedures.

A central result in this literature is the incompatibility of multiple fairness criteria under realistic conditions, implying that no single metric can satisfy all normative desiderata simultaneously [12]. Consequently, algorithmic fairness frameworks require external normative choices regarding which constraints to prioritize, leaving unresolved how institutions should structure decisions before outcomes are realized.

Institutional applications of such tools have encountered similar challenges. As noted in **Section 1.2**, the College Board's Adversity Score was withdrawn due to opacity concerns, illustrating that statistical correction alone does not substitute for transparent, rule-based decision architecture. While algorithmic fairness provides valuable diagnostic tools, it offers limited guidance on how fairness should enter the admissions process at the decision level.

### 2.3. Socioeconomic Evidence and the Calibration Problem

A substantial body of research in education economics documents a robust association between socioeconomic background and educational outcomes. Family income, parental education, and neighborhood conditions explain a meaningful share of achievement variance across cohorts and countries [13,14]. Importantly, this relationship is probabilistic rather than deterministic, indicating that disadvantage shapes opportunity without fixing individual potential.

Empirical work documents steep, continuous gradients between socioeconomic background and achievement/mobility, suggesting that fairness mechanisms should be sensitive to these continuous patterns of disadvantage [15,16]. Related evidence on "Lost Einsteins" further documents that high-ability individuals from low-income backgrounds are disproportionately absent at elite thresholds [17], reinforcing the need for calibrated, marginal interventions rather than categorical adjustments. However, despite the strength of this evidence base, the literature remains largely descriptive. It identifies inequality patterns and mobility gaps but does not specify how such findings should be translated into operational decision rules that preserve merit-based standards.

As a result, socioeconomic evidence supports the need for calibrated correction but does not itself resolve how much adjustments should be applied, at what stage of the decision process, or under what institutional constraints. This gap highlights a calibration problem: empirical insights alone are insufficient to guide executable admissions design.

### 2.4. Policy Implementation, Strategic Response, and Procedural Constraints

Fairness mechanisms are implemented within strategic institutional environments where actors respond to incentives and rules. Universities, administrators, and applicants adapt to policy structures, often producing unintended consequences that undermine stated equity goals [18]. Zero-sum allocation frameworks are particularly vulnerable to such dynamics, as they intensify competition and encourage strategic behavior.

Policy research on procedural justice emphasizes that public acceptance depends not only on distributive outcomes but also on the transparency, consistency, and auditability of decision processes [19,20]. Systems that rely on discretionary adjustments or opaque criteria risk eroding legitimacy, even when motivated by equity considerations.

These insights point to a final design constraint: effective fairness mechanisms must constrain discretion through rule-based execution while remaining compatible with institutional autonomy and capacity limits. Without a clearly specified decision sequence, equity interventions remain susceptible to gaming, political contestation, and administrative drift.

Taken together, existing literatures identify persistent constraints—proxy imprecision, ex post adjustment, calibration ambiguity, and strategic vulnerability—without offering an integrated decision architecture (see **Appendix A** for supplementary context). The AMF responds to this gap by treating fairness as a decision-level design problem, specifying where and how socioeconomic correction enters the decision sequence under fixed procedural rules. **Section 3** formalizes this mechanism.

## 3. Methodology: An Adaptive Merit Framework

This section formalizes the Adaptive Merit Framework (AMF) introduced earlier and specifies the correction rule and implementation used in the analysis. **Fig. 2** provides an overview of the algorithmic execution flow of AMF, which is formalized in the subsections below.

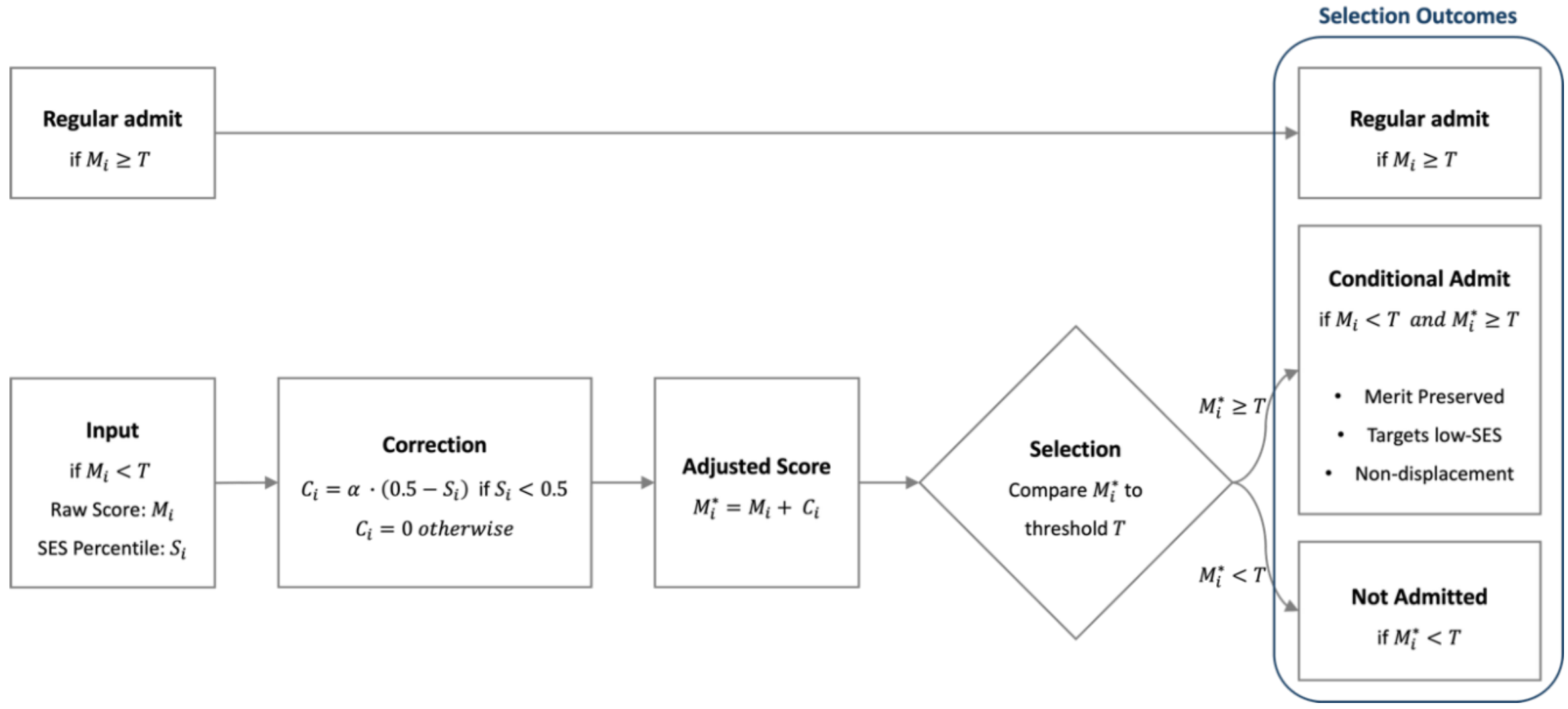

**Figure 2. Algorithmic execution flow of the Adaptive Merit Framework (AMF),** illustrating non-displacing conditional selection relative to a fixed raw-score threshold.

### 3.1. Dual-Selection Structure and Correction Rule

The Adaptive Merit Framework (AMF) adjusts each applicant's raw performance score using a linear SES-based correction. However, before defining the adjusted score, it is important to clarify the dual-selection structure of AMF. Regular admissions are determined exclusively by raw performance scores and remain unchanged. SES-based corrections are applied solely to assess eligibility for additional, non-displacing admits, ensuring that no regular admit is displaced by the corrections.

Regular admissions are determined as:

$$M_i \geq T \tag{1}$$

where, $T$ is the raw-score threshold corresponding to the top-$k$ positions. In the empirical analysis, the threshold T is operationalized as the score corresponding to the top 10% of applicants within each cohort. This choice serves as a simulation benchmark and does not affect the structure of AMF, which treats the threshold as exogenous.

The general specification of the correction rule is:

$$C_i = \alpha \cdot (\mu - S_i) \tag{2}$$

where, $S_i \in [0,1]$ is the normalized SES index and $\mu$ denotes the population mean of $S_i$.

This formulation preserves generality: when SES is measured on an arbitrary non-percentile scale, $\mu$ reflects the empirical center of the distribution and determines which portion of applicants are eligible for correction.

In this study, SES is percentile-normalized, yielding an approximately uniform distribution on [0,1] in large samples. Consequently, its population mean is expected to be close to 0.5:

$$\mu = E[S_i] = 0.5 \tag{3}$$

Substituting this value into the general formulation yields the empirical correction rule used in the PISA-based analysis:

$$C_i = \alpha \cdot (0.5 - S_i) \tag{4}$$

Thus, applicants with $S_i < 0.5$ receive a positive correction, while those at or above the empirical median receive none. This property is not a design choice but an immediate consequence of percentile normalization, which implies that AMF focuses exclusively on structurally disadvantaged applicants.

Corrected performance is computed as:

$$M_i^* = M_i + C_i \tag{5}$$

and applicants qualify as additional admits when:

$$M_i^* \geq T \tag{6}$$

Because thresholds are determined solely from raw scores, AMF expands opportunity **without displacing regular admits.** (derivation details in **Appendix C**)

### 3.2. SES Measurement and Normalization

SES is measured using PISA's ESCS composite index. To prevent distortion of the percentile transformation, extreme outliers are removed using a $1.5 \times IQR$ rule (14 cases removed out of 6,377). Let $ESCS_i$ denote the cleaned composite socioeconomic index. The SES index used for correction is defined as:

$$S_i = PercentileRank(ESCS_i) \tag{7}$$

This transformation maps socioeconomic status onto a unit interval while preserving individuals' relative positions in the population distribution.

In the PISA 2022 Korea dataset, the applicant pool represents a nationally representative sample of 15-year-olds. Accordingly, SES percentiles reflect population-level rather than self-selected distributions. In institutional implementations, SES percentiles should be computed relative to the national population. If percentiles were instead calculated within a self-selected applicant pool, the empirical mean would shift upward, systematically under-correcting genuine disadvantage and weakening the intended targeting of the correction function.

### 3.3. Empirical Calibration of $\alpha$

To ensure that AMF remains empirically grounded and conservative in scale, the policy parameter $\alpha$ is calibrated using the observed SES-achievement gradient in the PISA 2022 Korea dataset. Prior analyses show that a one-standard-deviation increase in socioeconomic status corresponds to an achievement gap on the order of several dozen score points, providing an empirical benchmark for the magnitude of structural disadvantage.

Accordingly, we select values of $\alpha$ that represent modest fractions of this gradient ($\alpha \in \{5, 10, 15\}$), ensuring that corrections recover suppressed potential without materially reshaping the overall performance distribution. Across these values, the resulting adjustments remain small relative to competitive thresholds and do not elevate low-performing applicants above merit-based cutoffs.

Full estimation details for the calibration procedure are provided in **Appendix B.2**.

### 3.4. Simulation Protocol Using PISA 2022 Korea Data

All empirical analyses use the 6,377 Korean examinees in PISA 2022 after outlier removal. For each value of the policy parameter $\alpha \in \{5, 10, 15\}$, admissions are simulated following the algorithmic execution flow of AMF (**Fig. 2**), defined in **Section 3**.

Regular admissions are determined exclusively by raw scores, while SES-based corrections are applied only to evaluate eligibility for additional, non-displacing admits. Corrected scores are computed accordingly, and conditional admits are identified relative to the same raw-score threshold.

The analysis examines (i) SES distribution of admits, threshold gaps, and correction magnitude; (ii) robustness to SES noise, score perturbations, and threshold shifts; and (iii) long-run dynamics using a Dynamic Bayesian Network (DBN) model.

Detailed derivations, implementation procedures, and robustness pipelines are provided in the **Appendices C** and **D**.

## 4. Results

This section presents the empirical behavior of the Adaptive Merit Framework (AMF) when applied to the PISA 2022 Korea mathematics dataset ($N$ = 6,377). **Fig. 3** provides a comprehensive visualization of the key findings: the number of additional admits, their SES composition, and the magnitude of threshold gaps. All results follow the implementation specified in **Section 3** and the simulation protocol detailed in **Appendix D**.

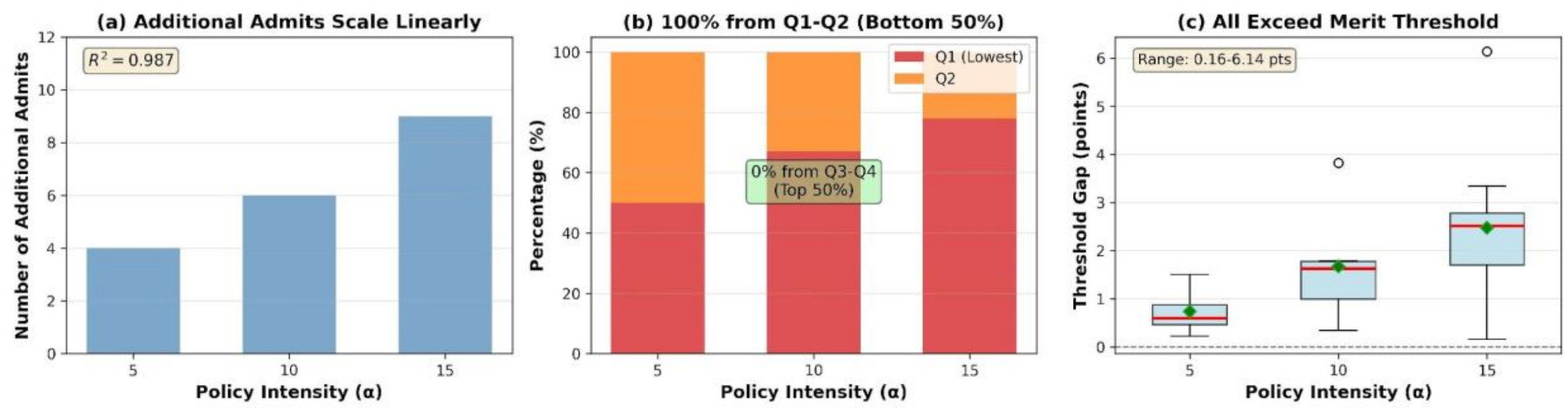

**Figure 3. Empirical Results from PISA 2022 Korea (*N* = 6,377)**. (a) Additional admits scale linearly with $\alpha$ ($R^2$ = 0.987). (b) 100% originate from bottom 50% SES under baseline conditions (Q1-Q2). (c) All exceed threshold T=666.62 by 0.16-6.14 points.

### 4.1. Baseline Performance and Merit Threshold

The raw top 10% threshold for the sample is $\mathrm{T} = 666.62$. The 638 students above this cutoff constitute the **regular admits** with **Eq. (1)**. No part of AMF modifies this threshold or displaces students who exceed it.

### 4.2. Correction Magnitudes

Using percentile-normalized SES and the correction rule **Eq. (4)**, AMF produces bounded and moderate adjustments across $\boldsymbol{\alpha} \in \{\mathbf{5}, \mathbf{10}, \mathbf{15}\}$, conditional on eligibility for additional admission.

**Table 1. Correction Magnitudes Among Conditional Admits (PISA 2022 Korea).**

| $\alpha$ | **Min $C_i$** | **Max $C_i$** | **Mean $C_i$** |
|---|---|---|---|
| **5** | 0.99 | 2.32 | 1.48 |
| **10** | 1.98 | 4.64 | 2.99 |
| **15** | 2.97 | 6.95 | 4.76 |

**Table 1** summarizes the magnitude of SES-based corrections among conditional admits across policy intensities**.** Across all values of α, correction magnitudes remain modest relative to the empirical SES-achievement gradient, indicating conservative, non-disruptive adjustments. Full correction distribution are reported in **Appendix B.2.**

### 4.3. Additional Admits Under AMF

For each value of $\alpha$, AMF identifies applicants whose corrected performance exceeds the existing merit threshold, while their original raw scores remain below it (see **Eqs**. $(\mathbf{5}) - (\mathbf{6})$). The resulting additional admits are summarized in **Table 2**.

**Table 2. Number of Additional Admits by Policy Intensity $\boldsymbol{\alpha}$.** (PISA 2022 Korea, N=6,377, Threshold T=666.62)

| $\alpha$ | **Additional admits** | **Share of cohort** | **Mean $C_i$** |
|---|---|---|---|
| **5** | 4 | 0.06% | ~1.48 |
| **10** | 6 | 0.09% | ~2.99 |
| **15** | 9 | 0.14% | ~4.76 |

*Note*: Population-weighted estimates following OECD sampling weights are reported in **Appendix D.4** and exhibit the same directional pattern.

Across all settings, AMF expands admissions by fewer than 0.15% of the cohort. As shown in **Fig. 3(a)**, under $\alpha$ = 5, AMF identifies 4 additional admits (0.06% of the cohort). Under $\alpha$ = 10, this increases to 6 students (0.09%). Under $\alpha$ = 15, 9 students qualify (0.14%). The linear scaling pattern implies the predictability and transparency of the mechanism.

**Fig. 3(c)** shows that all conditional admits exceed the merit threshold by 0.16 to 6.14 points, implying that corrections recognize suppressed performance rather than relaxing standards. Detailed score profiles and threshold-gap values appear in **Appendix D.2**.

### 4.4. SES Composition of Conditional Admits

A central goal of AMF is to identify students whose raw performance is understated due to structural disadvantage. The SES distribution of conditional admits shows clear evidence of this targeting.

**Table 3. SES Quartile Composition of Conditional Admits.** All 100% from Q1-Q2 across all *alpha* values. No students from Q3-Q4 qualified under any parameter setting.

| $\alpha$ | **Q1** | **Q2** | **Q3** | **Q4** |
|---|---|---|---|---|
| **5** | 50% | 50% | 0% | 0% |
| **10** | 67% | 33% | 0% | 0% |
| **15** | 78% | 22% | 0% | 0% |

Across all $\alpha$ values, conditional admits originate exclusively from the bottom half of the SES distribution, with no students from Q3-Q4 qualifying under any parameter setting.

This pattern is also reflected in **Fig. 3(b)** and remains stable under robustness checks (**Appendix D.3**; see **Fig. D.2**), with only minor boundary effects under SES measurement noise.

### 4.5. Interpretation

Three empirical features characterize AMF in the Korean PISA sample.

First, AMF induces a selective expansion: the number of additional admits remains small (4-9 students), consistent with prior evidence that high-achieving but structurally disadvantaged students constitute a rare population in large cohorts.

Second, this expansion is tightly targeted. Conditional admits originate predominantly from the lowest SES quartiles (Q1-Q2), indicating that AMF responds to structural disadvantage rather than proximity to SES neutrality.

Third, all conditional admits exceed the original merit threshold after correction, consistent with the pattern observed in **Section 4.3**.

Taken together, these results suggest that modest, transparent corrections can expand opportunity in a conservative and institutionally stable manner.

## 5. Decision Spine of the Adaptive Merit Framework

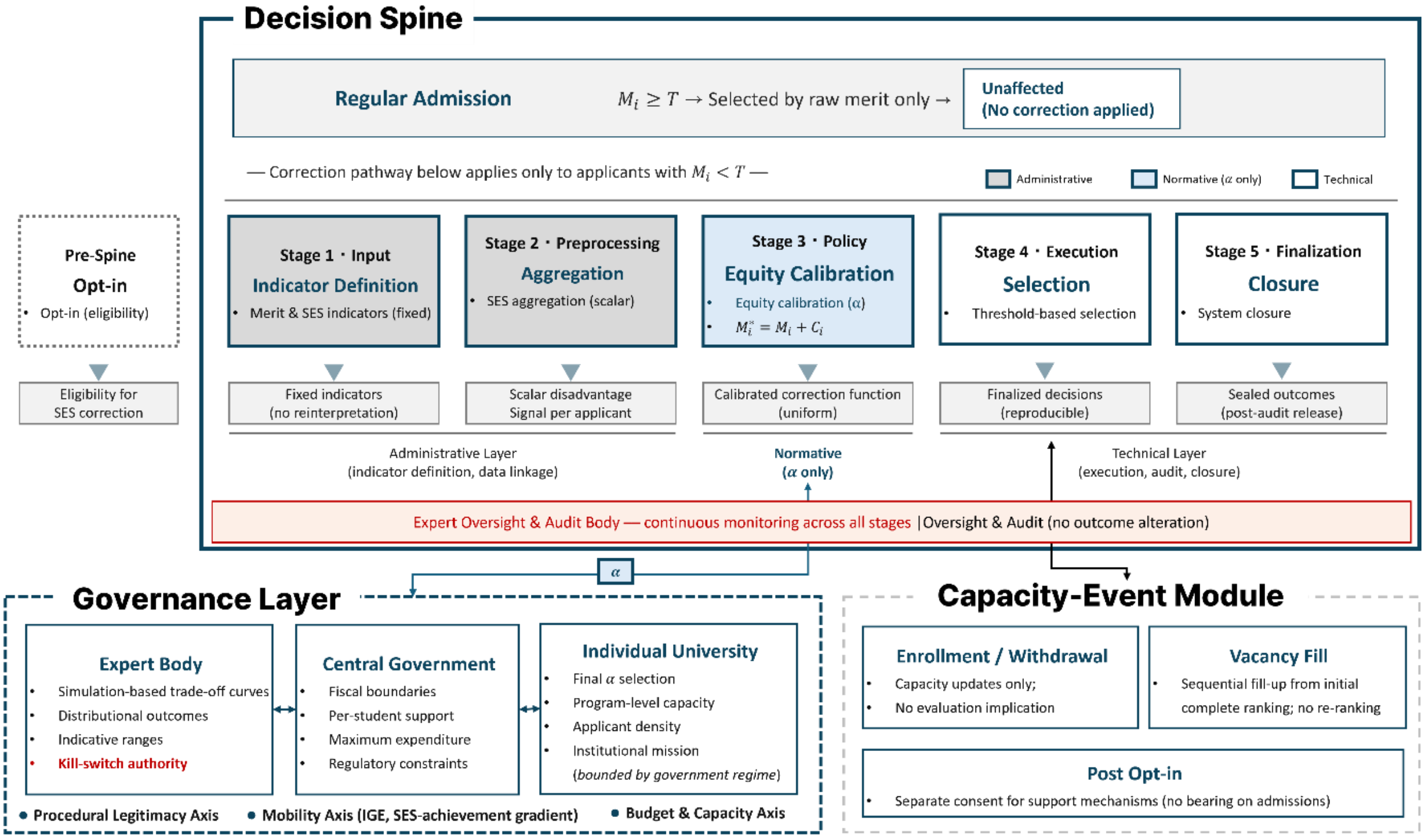

**Figure 4. Decision spine of the Adaptive Merit Framework (AMF).** The upper path shows that regular admissions ( $\boldsymbol{M_i} \geq \boldsymbol{T}$ ) remain untouched. The five-stage spine below governs the correction pathway for conditional admits. **Stage 3** (highlighted) is the sole normative intervention point. The tripartite governance layer determines through three calibration axes. The expert oversight body monitors the entire spine with kill-switch authority.

Admission decisions are not isolated choices, but structured processes composed of multiple institutional commitments. This section defines the decision spine of the Adaptive Merit Framework (AMF): an invariant sequence of decision stages through which all AMF-based implementations must pass. While implementation details may vary across institutional contexts, the ordering and functional role of each stage remain fixed.

The decision spine specifies where decisions are made, where discretion is deliberately removed, and how outcomes are executed and closed. In doing so, it transforms admissions from a collection of discretionary judgments into a reproducible decision pipeline.

AMF explicitly separates three layers:

- Normative layer: policy intent, represented exclusively by $\alpha$
- Administrative layer: indicator definition, data linkage, eligibility verifications
- Technical layer: aggregation, weighting, and execution

Only the first layer is normatively adjustable. The latter two layers are constrained to be auditable, non-discretionary, and non-ranking-altering.

From a planning perspective, AMF can be interpreted as a structured decision problem in which institutional actors select a limited set of configurable levers (indicator sets, aggregation rules, the equity calibration parameter α), subject to institutional capacity constraints.

### 5.1. Structural Definition of the Decision Spine

The AMF decision spine consists of five ordered stages:

(1) input definition,

(2) aggregation and weighting,

(3) equity calibration,

(4) execution, and

(5) closure.

Each stage performs a distinct function and produces outputs that serve as inputs to the next stage. The sequence is non-substitutable: removing or reordering any stage alters the nature of the decision process itself. Decision making in AMF is thus distributed across the entire spine rather than concentrated at a single point.

This structural commitment prevents decisions from being deferred, relocated, or retrospectively modified. Once instantiated, the spine governs all downstream outcomes. In contrast to fragmented equity tracks or ad hoc adjustments, the decision spine consolidates all policy levers into identifiable stages, enabling ex ante analysis of trade-offs and ex post auditability.

**Pre-Spine Boundary: Opt-in to Data Disclosure.**

Prior to entering the decision spine, applicants may opt in to the disclosure of socioeconomic information used for disadvantage measurement. This opt-in determines eligibility for SES-based correction but does not guarantee any adjustment or admission outcome. Once exercised, the choice is binding for the execution cycle and cannot be revised ex post.

### 5.2. Input Decision: Indicator Definition and Data Instantiation

**Input.**

For each applicant $i$, the decision spine requires a fixed set of inputs, including a merit score and a measure of socioeconomic disadvantage. The selection of disadvantage indicators constitutes an explicit policy decision, as it determines which dimensions of inequality are recognized by the system.

Institutions may generate the merit input using any internally defined evaluation structure, including multi-metric composite systems. The AMF requires only the final scalar merit signal $M_i$, without imposing constraints on how institutions construct it.

**Processing.**

In the empirical analysis, socioeconomic disadvantage is represented using the PISA ESCS index, a self-reported proxy adopted to enable to population-level simulation. In operational settings, however, indicator selection is a normative institutional decision. The same stage may instead be instantiated using administratively verified income and asset records, parental education, regional indicators, or composite measures derived from multiple sources.

To be admissible within the decision spine, disadvantage indicators must satisfy four properties:

(1) a continuous, percentile-based structure to enable fine-grained and distributionally stable correction;

(2) reliance on verified administrative data or validated composite measures;

(3) temporal stability sufficient to prevent short-term manipulation or volatility; and

(4) privacy-compliant data handling.

In most operational contexts, the fourth requirement is naturally satisfied when disadvantage indicators are derived from legally collected administrative records and applied only to applicants who have provided explicit consent through pre-spine opt-in procedures. Under such conditions, the use of socioeconomic data remains both procedurally legitimate and institutionally compliant.

**Output.**

Once defined, the disadvantage indicator is fixed and passed unchanged to the aggregation stage. No subsequent adjustment or reinterpretation of inputs is permitted within the spine.

**Modularity.**

While the specific indicators employed are modular and context-dependent, the requirement that inputs be explicitly defined prior to processing is invariant.

### 5.3. Aggregation and Weighting: Non-discretionary Technical Processing

**Input.**

The aggregation stage receives the fixed disadvantage indicators defined in the previous step, potentially comprising multiple correlated dimensions. It determines how these indicators are transformed before entering the allocation rule.

**Processing.**

These inputs are transformed into a single disadvantage signal using a pre-specified statistical procedure. In the empirical analysis, disadvantage is proxied by PISA's ESCS (a pre-constructed composite index). In operational settings where multiple administrative indicators are available, institutions may adopt a rule-based aggregation method (e.g., PCA) fixed ex ante.

Aggregation is intentionally delegated to a technical procedure. Once fixed, neither weights nor aggregation outcomes can be modified on a case-by-case basis, thereby eliminating discretionary intervention.

**Output.**

The aggregation process produces a scalar disadvantage signal for each applicant, which is then used mechanically in the correction rule.

**Modularity.**

Alternative aggregation methods may be substituted depending on data characteristics or institutional preference, provided that the aggregation rule is fixed prior to execution and insulated from discretionary intervention.

In practice, administratively valid disadvantage indicators largely collapse into socioeconomic dimensions, even when multiple variables are used. This reflects institutional constraints rather than conceptual limitation: indicators eligible for rule-based correction must be persistent, generalizable, and ex ante verifiable. As a result, most correctable disadvantage operates within the broad category of environmental and socioeconomic conditions.

At this point, aggregation concerns only the construction of disadvantage signals. The scope of correction, however, raises a separate normative question. While the correction function can, in principle, be applied to any evaluative dimension, the determination of legitimate correction targets lies outside the decision spine. In this study, correction is applied to academic merit, while extension to other evaluative dimensions is treated as a policy choice external to the decision spine.

## 5.4. Equity Calibration as a Policy Lever ($\alpha$)

### 5.4.1. Conceptual Structure and Layer Separation

#### (1) Merit Construct vs. SES Signal: Different Types of Inputs

Institutions may evaluate applicants using multiple performance indicators. These indicators—such as exam scores, coursework, interviews, or structured assessments—are institution-defined measures of *achievement*. Their aggregation defines the institutional merit construct:

$$M_i = f(m_{i1}, m_{i2}, \dots, m_{in}) \tag{8}$$

where, $m_{ij}$ denotes the *j-th* indicator for applicant $i$, and $f(\cdot)$ is any legitimate composite or weighting rule chosen by the institution. This formulation preserves institutional autonomy and places no restriction on the internal structure of merit evaluation.

By contrast, the SES variable $S_i$ is not an additional merit indicator. It is an *opportunity signal*, representing structural conditions under which achievement is realized. Its statistical nature is also different: commonly, SES is already an aggregate of socioeconomic background components (e.g., parental education, occupation, income proxies) and functions as a single contextual measure rather than a performance metric.

**(2) Why SES Must Not be Embedded Inside the Merit Function**

Embedding SES within $f(\cdot)$ would collapse the conceptual boundary between achievement and opportunity. SES would become indistinguishable from performance indicators, eliminating its role as a **policy lever** and making equity adjustments non-transparent.

Moreover, mixing SES into the merit function complicates interpretation, varies across institutions, and undermines the stability required for cutoff anchoring and vacancy-fill procedures.

Thus, SES must operate at a **structurally separate layer**, not as a competing term inside the institutional merit construct.

**(3) The AMF Correction Layer**

AMF preserves the institutional merit construct in full and applies SES-based calibration only **once**, after merit has been instantiated (**Eqs. (4)-(5)**):

$$M_i^* = M_i + \alpha \cdot (0.5 - S_i)$$

This separation maintains:

1. **Transparency**: The institutional merit construct remains untouched.

2. **Modularity**: SES correction functions as an independent add-on module.

3. **Policy Leverage**: The equity parameter $\alpha$ is directly interpretable and adjustable without altering institutional scoring schemes.

4. **Comparability**: The correction applies consistently across heterogeneous institutional merit constructs.

**(4) Implication for the Decision Spine**

This layered design aligns directly with the decision spine:

- **Input Layer**: institutions instantiate $M_i$ in any form they choose.
- **Equity Layer**: AMF applies opportunity correction separately.
- **Execution Layer**: vacancy fill / cutoff anchoring operate on the corrected signal $M_i$.

The separation between achievement and opportunity signals is therefore not cosmetic—it is the structural basis that ensures AMF behaves as an interpretable, stable, and policy-relevant mechanism.

### 5.4.2. Operational Calibration and Governance of $\alpha$

**Input.**

The equity calibration stage receives aggregated disadvantage signals and the merit-anchored decision rule specified earlier in the framework.

**Processing.**

Normative intervention is confined to a single parameter, $\alpha$, which governs the strength of socioeconomic correction applied within the decision rule. Adjusting $\alpha$ does not alter the ordering of applicants by merit or the structure of the rule itself; rather, it modulates the magnitude of correction in a transparent and continuous manner.

Importantly, $\alpha$ does not determine admission outcomes directly. Instead, it governs the sensitivity of score adjustment, whose realized effects depend on local score distributions and competitive density. As a result, identical values of $\alpha$ may lead to different admission outcomes across institutions, while preserving the same underlying decision logic.

The determination of $\alpha$ is explicitly political rather than technical. Analytical tools—such as sensitivity analysis, simulation-based trade-off curves, or budget constraints—may inform its selection, but they do not determine it.

**Output.**

The output of this stage is a calibrated correction function that maps disadvantage signals into score adjustments, applied uniformly across all applicants.

**Modularity.**

While the procedure used to select $\alpha$ is modular (e.g., mobility targets, fiscal limits, or scenario analysis), the role of $\alpha$ as the sole normative policy lever is fixed within the decision spine.

**Operational Governance and Calibration Criteria**

In operational settings, the choice of $\alpha$ is embedded in **a tripartite governance structure** involving (1) an expert oversight and audit body, (2) the central government, and (3) individual universities.

First, the expert body generates simulation-based trade-off curves that map alternative values of $\alpha$ to distributional outcomes, social mobility indicators, and indicative ranges of additional admits under representative or population-level score distributions. These simulations are designed to clarify structural trade-offs rather than to produce institution-specific forecasts.

Second, the central government defines fiscal and regulatory boundaries—such as per-student support levels of maximum allowable expenditure—within which $\alpha$ may be selected. Rather than fixing $\alpha$ directly, public authorities bound the feasible space of implementation through budgetary and regulatory commitments.

Finally, within this feasible range, individual universities act as final decision makers, selecting an $\alpha$ that aligns with their program-level capacity, applicant density, and institutional mission, based on their own score distributions and operational constraints.

Conceptually, $\alpha$ is calibrated along **three decision axes**.

First, a **procedural legitimacy axis** requires that $\alpha$ be set through a transparent and periodically reviewed rule-making process informed by published simulations rather than ad hoc political pressure.

Second, a **mobility axis** links $\alpha$ to structural indicators such as intergenerational income elasticity and the SES-achievement gradient: steeper gradients justify higher values of $\alpha$, while flatter gradients permit gradual reduction over time.

Third, a **budget and capacity axis** constrains the upper bound of $\alpha$ using estimates of indicative additional admits, per-student costs, and institutional seat capacity, ruling out settings that would violate fiscal or infrastructural limits.

Beyond generating simulation-based trade-off curves, the expert oversight body continuously monitors the procedural integrity of the entire calibration process. It conducts real-time audits of data logs at each stage and is granted a temporary suspension (kill-switch) authority in cases of procedural violations, such as parameter values exceeding predefined bounds, inconsistencies between input data and outputs, or indications of strategic manipulation favoring specific groups. This authority is strictly limited to safeguarding system integrity and does not extend to altering outcomes or recalibrating policy parameters. To prevent strategic gaming and political interference, the expert oversight body operates alongside the decision spine, ensuring that calibration procedures and administrative data verification remain tamper-proof.

**Optional public observability of $\alpha$.**

Although $\alpha$ is defined as a single normative policy lever within the decision spine, it may be made publicly observable ex ante, together with simulation-based mappings between alternative values of $\alpha$, indicative public investment levels, and institutional selections. Such observability reduces informational asymmetries without creating additional individual-level strategic margins.

### 5.5. Execution and Closure: Batch Processing and Irreversibility

**Input.**

The execution stage receives fully specified decision rules, calibrated correction functions, and applicant data processed through all prior stages.

**Processing.**

Admissions decisions are implemented through batch execution, applying identical rules to all applicants simultaneously. Eligibility for AMF-based processing may be governed by opt-in mechanisms that define procedural boundaries without introducing individualized evaluation.

Once execution begins, individual-level exceptions and post hoc adjustments are structurally excluded. Governance boundaries are enforced to prevent retrospective modification of outcomes.

**Output.**

The output of this stage is a finalized set of admission decisions, reproducible given identical inputs and parameters.

Where institutional capacity permits, the same information used for admission may be propagated to post-admission support mechanisms, such as financial aid or academic assistance. This preserves continuity between selection and support while maintaining irreversibility of admission outcomes.

**Modularity.**

The form of opt-in mechanisms, execution cycles (e.g., annual or multi-cycle), post-admission support structures, and audit or appeal procedures may vary by institutional context. Their functional rule—ensuring batch processing and closure—remains invariant.

After batch execution is completed, an independent expert oversight and audit body performs a consistency check between the input and the finalized admission outcomes. This verification is conducted using anonymized applicant identifiers, without access to personally identifiable information. Only upon confirmation that no discrepancies or procedural violations are detected is the system formally sealed (closure) and the outcomes released. This final audit step reinforces irreversibility by ensuring that execution integrity is verified before results become effective.

### 5.6. Capacity-Event Module (Execution-Level Events)

#### Enrollment/Withdrawal

Capacity updates only; no implication for evaluation.

#### Post Opt-in

Separate opt-in for support mechanisms; no bearing on admissions. Participation in post-admission support mechanisms is governed by a separate opt-in decision exercised after admission outcomes are finalized. This choice affects only the use of information for support allocation and has no bearing on admission decisions.

**Vacancy Fill**

Conducted via sequential fill-up using the initial complete ranking; consistent with global admission practices; ensures path-independence and prevents re-ranking.

Waitlist admissions are treated as an execution-phase extension rather than a new decision problem. Once the AMF ranking is fixed, subsequent offers follow the established order without recalibration, ensuring procedural consistency and avoiding ex-post discretion.

### 5.7. Summary

The decision spine of AMF specifies a complete and ordered decision pipeline, from input definition to irreversible execution. While individual modules may be instantiated differently across institutional contexts, the spine itself defines the minimal structure required to translate equity objectives into operational admissions decisions without discretionary fragmentation.

In this sense, the decision spine functions as a minimal decision-making framework: it converts high-level equity commitments into a sequence of operational choices that can be planned, simulated, and audited ex ante.

## 6. Discussion

This study proposed and empirically evaluated the AMF as a procedural mechanism for incorporating socioeconomic disadvantage into merit-based admissions under a non-displacement constraint. By combining a fixed decision spine with a single policy calibration parameter, AMF operationalizes fairness as an ex-ante design problem rather than an ex-post adjustment. The results from the PISA 2022 Korea sample clarify what such a design achieves—and, equally importantly, what it deliberately does not.

### 6.1. Selective Expansion and the Scarcity of Suppressed Merit

A central empirical finding is that AMF generates only a small number of additional admits across all policy intensities tested. Even under the most permissive calibration ($\alpha = 15$), fewer than 0.15% of the cohort qualifies as conditional admits. This scarcity is not an artifact of conservative parameter choices, but a structural feature of the mechanism operating near an existing merit threshold.

This pattern aligns with prior evidence that high-achieving students from disadvantaged backgrounds constitute a rare population in large cohorts. Rather than revealing a large pool of overlooked candidates, AMF identifies a small number of applicants whose observed performance lies just below competitive cutoffs and whose scores are plausibly suppressed by structural disadvantage. In this sense, the mechanism does not manufacture opportunity at scale; it recovers marginal cases that standard merit signals fail to recognize.

Importantly, the linear scaling of additional admits with $\alpha$ indicates predictability rather than volatility. Policymakers can anticipate the magnitude of expansion associated with a given calibration, avoiding the uncertainty that often accompanies quota-based or categorical interventions. Moreover, the governance framework specified in **Section 5.4.2** structurally constrains the feasible range of $\alpha$ through empirical calibration principles, budget-linked feasibility conditions, and independent oversight, ensuring that institutional autonomy over $\alpha$ operates within analytically grounded and fiscally bounded limits rather than as unconstrained discretion.

### 6.2. Consistent SES Targeting Without Proxy Leakage

Across all parameter settings, conditional admits originate exclusively from the bottom half of the SES distribution, with no representation from the top two quartiles. This outcome follows directly from the percentile-based correction rule and confirms that the mechanism targets structural disadvantage rather than proximity to SES neutrality.

The concentration of admits in Q1 and Q2 demonstrates that AMF avoids a common failure mode of proxy-based equity policies, where disadvantaged individuals may benefit indirectly through coarse group classifications. Here, the correction operates continuously and monotonically with respect to SES, preventing threshold-crossing advantages for higher-SES applicants by construction.

Robustness analyses further indicate that this targeting remains stable under reasonable perturbations to SES measurement and score variance. Minor boundary effects appear only under extreme noise assumptions, suggesting that the observed composition is not fragile to realistic data imperfections.

### 6.3. Merit Preservation and Non-Displacement by Construction

A defining feature of AMF is that all regular admits—those exceeding the raw merit threshold—remain unaffected by the correction process. Conditional admission occurs only when an applicant's corrected score exceeds the same threshold while the raw score does not. As a result, the mechanism expands opportunity without displacing any existing admit or relaxing standards.

Empirically, all conditional admits exceed the merit cutoff by a positive margin after correction, with threshold gaps ranging from small but nontrivial values to several points. This pattern reinforces that AMF recognizes suppressed performance rather than compensating for low ability. The mechanism does not elevate weak performers above competitive standards; it adjusts for environmental disadvantages that plausibly depress observed scores near the margin.

This distinction is critical for institutional legitimacy. By preserving the original threshold and avoiding rank reordering among regular admits, AMF maintains a transparent link between merit and selection outcomes.

### 6.4. Interpreting Correction Magnitudes and Policy Intensity

The magnitude of SES-based corrections among conditional admits remains modest across all calibrations. Mean corrections range from approximately 1.5 to 4.8 points, with upper bounds well below the empirically estimated SES-achievement gradient. These adjustments are large enough to alter marginal outcomes but too small to reshape the overall performance distribution.

This boundedness highlights the role of $\alpha$ as a policy intensity parameter rather than a redistributive lever. Adjusting $\alpha$ scales the strength of correction smoothly, allowing institutions to select conservative or slightly more permissive regimes without introducing discontinuities or categorical breaks. In contrast to quota systems, where small parameter changes can induce large reallocations, AMF's calibration behaves proportionally and predictably.

### 6.5. System-Level and Institutional Considerations

Although the empirical analysis is conducted at the cohort level, AMF is designed to operate within broader institutional systems. Because the correction rule and selection logic are fixed, implementation does not require discretionary judgements by admissions officers once parameters are set. This separation between rule design and execution supports auditability and reduces the risk of ad hoc interference.

At the system level, anchoring SES percentiles to population-wide distributions is essential. Computing disadvantage relative to self-selected applicant pools would systematically bias corrections downward and undermine the intended targeting. The framework therefore presumes integration with administrative data infrastructures capable of producing verifiable, population-referenced SES measures.

From a dynamic perspective, AMF's modest short-run adjustments are compatible with long-run modeling of educational trajectories. Small early-stage corrections may accumulate through subsequent transitions without destabilizing selection systems, suggesting potential complementarity with dynamic policy analysis rather than conflict.

AMF operates as an ex-ante mechanism, expanding opportunity at the point of selection rather than compensating through ex-post transfers. Ex-ante correction and post-admission support (e.g., financial aid, tutoring) are complements, not substitutes: the former ensures access, the latter supports persistence. An integrated admission-support pipeline is formalized in **Appendix F.6**.

### 6.6. Ethical and Governance Implications

Ethically, AMF occupies a middle ground between formal equality and outcome equalization. It does not claim to equalize opportunity fully, nor does it rely on group-based preferences. Instead, it encodes a limited corrective response to structural disadvantage within an otherwise merit-based procedure.

This design mitigates several ethical concerns commonly raised against affirmative action mechanisms. The absence of displacement reduces zero-sum perceptions, while the transparency of the correction rule limits arbitrariness. At the same time, the framework makes explicit that fairness is not a natural property of scores, but a choice embedded in institutional design.

Governance-wise, concentrating normative discretion in a single parameter clarifies accountability. Rather than obscuring value judgments within complex scoring formulas or opaque holistic reviews, AMF makes the intensity of correction visible and contestable at the policy level.

### 6.7. Limitations and Scope

The analysis relies on a single national dataset and focuses on mathematics performance at age 15. While the structure of AMF is general, empirical magnitudes may differ across subjects, cohorts, and institutional contexts. Moreover, SES is operationalized through a composite index that, while administratively practical, cannot capture all dimensions of disadvantage.

These limitations do not undermine the core contribution of the framework but define its scope. AMF is not proposed as a universal solution to educational inequality, but as a design template for incorporating disadvantage into selection procedures under realistic constraints.

While the current implementation uses SES as the sole disadvantage signal, the correction function can in principle accommodate multiple dimensions (e.g., regional isolation, caregiving burden) through composite indices, provided each satisfies the eligibility criteria in **Section 5.2**. Such extensions raise calibration challenges—including double-counting among correlated signals—and represent a natural direction for future work.

## 7. Conclusion

### 7.1. Toward a Unified and Multi-Dimensional Framework

While this study focuses on SES-based corrections to test scores, the broader vision underlying AMF is a fundamental redesign of admissions systems toward two complementary goals:

**Track consolidation.** AMF replaces fragmented categorical quotas—such as regional balance or income-based tracks—with a unified framework in which all applicants are evaluated within a single corrected pool. This consolidation preserves seat counts while reducing administrative complexity and labeling effects (**Appendix F.1.1**).

**Multi-Dimensional Evaluation.** While the current focus is on test scores, the correction rule naturally extends to any composite merit index, enabling institutions to integrate non-cognitive indicators once measurable and validated evaluation systems are developed (**Appendix F.1**).

### Extensions Beyond Admissions

While this study addresses college admissions, AMF's design principles–individual measurement, dynamic threshold, transparency, and non-displacement structure–extend naturally to other selection contexts requiring merit-equity balance. Extensions may incorporate diverse forms of structural disadvantage (regional disparities, caregiving burdens, family disruption) within the same rule-based architecture, provided measurements remain valid and ethical. These applications require domain-specific validation but illustrate AMF's potential as a generalizable policy engineering framework beyond admissions.

### Summary

In sum, AMF demonstrates that fairness need not be a redistributive or adversarial choice. When implemented as a transparent, merit-anchored procedural rule, fairness becomes **additive**–expanding opportunity while maintaining institutional legitimacy. This perspective provides a foundation for future research and policy experimentation on practical and implementable pathways to enhance educational mobility.

Detailed implementation pathways, including data requirements, safeguards, and pilot strategies, are provided in **Appendix F**.

### Code and Data Availability

All code and data used in this study are publicly available

#### Code Repository

https://github.com/ava-jahlee/adaptive-merit-framework

- Core AMF implementation with robustness tests and DBN model (amf_engine.py)

- Main analysis script generating all simulation results (run_amf_all_results.py)

- Population-weighted estimation with visualization (amf_weighted_analysis.py)

- Requirements: Python 3.8+, pandas, numpy, matplotlib, scipy, seaborn

#### Data

PISA 2022 Korea: https://www.oecd.org.pisa/data/2022database/

Sample size: N = 6,377 (14 outliers removed via 1.5×IQR rule)

Variables used:

- PV1MATH (plausible values for mathematics performance)

- ESCS (Economic, Social and Cultural Status composite index)

**Reproducibility**

All results can be reproduced by running:

```
$ cd scripts/

$ python run_amf_all_results.py

$ python amf_weighted_analysis.py
```

This generates all tables, robustness checks, trajectory analyses, and population-weighted estimation presented in the paper. All reported results are generated using a fixed random seed. Re-running the simulations may lead to minor numerical variation but does not alter the qualitative patterns reported.

## Acknowledgments

This research did not receive any specific grant from funding agencies in the public, commercial, or not-for-profit sectors.

## Author Contributions

Jung-Ah Lee: Conceptualization, Methodology, Software, Formal analysis, Investigation, Data curation, Visualization, Writing – original draft, Writing – review and editing.

## Declaration of generative AI and AI-assisted technologies

The author used Claude (Anthropic) and ChatGPT (OpenAI) as AI assistants during code development and manuscript editing. All research design, analytical decisions, and intellectual contributions are solely the work of the author, who takes full responsibility for the content.

## Declaration of Competing Interests

The author declares no known competing financial interests or personal relationships that could have appeared to influence the work reported in this paper.

## Appendix A. Supplementary Context and Evidence on Proxy-Based Fairness

This appendix provides supplementary institutional, legal, and empirical context referenced in **Section 2**. These materials support the design constraints identified in the main text but are presented separately to preserve analytical clarity and concision in the core argument.

### A.1. Legal and Policy Context of Group-Based Admissions in the United States

Group-based affirmative action policies in U.S. higher education have been shaped by a series of landmark court decisions emphasizing individualized assessment over categorical preference. Beginning with *Regents of the University of California v. Bakke* [8], the Supreme Court prohibited rigid quota systems while allowing limited consideration of race as one factor among many. Subsequent rulings, including *Gratz v. Bollinger* [9], further constrained mechanical point-based systems, reinforcing the requirement that admissions decisions reflect holistic and individualized evaluation.

Most recently, *Students for Fair Admissions v. Harvard* [6] effectively prohibited the explicit use of race in admissions, underscoring the legal fragility of group-based proxies even in the presence of persistent racial inequality. These decisions collectively illustrate the increasing difficulty of sustaining categorical fairness mechanisms within legally permissible admissions frameworks.

This trajectory highlights a structural tension: while group-based interventions were initially designed to correct historical disadvantage, evolving legal standards have progressively narrowed the range of admissible proxy-based corrections, regardless of empirical evidence on inequality.

### A.2. Proxy-Based Admissions Policies Outside the United States: The Korean Case

Comparable challenges have emerged in non-U.S. contexts. In South Korea, regional balance admissions and special-track policies have been implemented to address geographic and socioeconomic disparities in access to elite universities. However, administrative data indicate that the metropolitan share of admitted students has continued to increase over time, despite the formal expansion of regional quotas [5].

These outcomes suggest that regional proxies insufficiently capture individual disadvantage, particularly as demographic composition, migration patterns, and applicant strategies adapt to policy rules. As with race-based proxies in the U.S. regional categories exhibit declining precision over time, limiting their effectiveness as stable fairness instruments.

### A.3. Extended Evidence on Socioeconomic Gradients in Educational Achievement

A substantial empirical literature documents the association between socioeconomic status and educational outcomes. Analysis of PISA microdata indicates that socioeconomic background explains approximately 15% of the variance in student performance across OECD countries, with correlation coefficients between SES indices and achievement scores commonly exceeding 0.35 [13,21].

National administrative data further reveal sharp attrition patterns along socioeconomic lines. For example, Korean longitudinal studies report that approximately 22% of students from low-income backgrounds exit competitive academic tracks between middle and high school, despite comparable early achievement [1].

Importantly, these patterns reflect probabilistic gradients rather than deterministic sorting. High-achieving students from disadvantaged backgrounds remain systematically underrepresented among top academic performers. Individuals with test scores in the top percentile disproportionately originate from upper-income households, with those from the bottom income quintile substantially underrepresented—a phenomenon often described as the "Lost Einsteins" problem [17].

### A.4. Illustrative Institutional Cases: Contextual Scoring and Adversity Indices

Several institutional attempts have sought to incorporate contextual disadvantage into admissions decisions. One prominent example is the College Board's Adversity Score, which aggregated neighborhood- and school-level indicators to provide admissions officers with contextual information. While conceptually aligned with individualized assessment, the tool faced criticism for opacity, limited interpretability, and unclear procedural role within admissions decisions, contributing to its discontinuation in 2019.

Similar challenges have been observed in other contextual scoring systems that lack explicit integration into decision rules. Without clearly defined threshold or execution procedures, such tools risk being perceived as discretionary overlays rather than rule-based mechanisms, reducing institutional trust and political sustainability.

### A.5 Summary Implications

The supplementary evidence presented here reinforces the constraints identified in **Section 2**. Group-based proxies face legal and precision limits, empirical SES gradients support calibrated rather than categorical correction, and institutional experiments with contextual scoring reveal the importance of transparent execution rules. These contextual materials motivate, but do not substitute for, the decision-level design formalized in the AMF.

## Appendix B. SES Measurement, Data Processing, and Normalization

### B.1. Construction of the SES Index (PISA 2022 Korea)

#### B.1.1. Raw SES Variable: ESCS

SES is operationalized using the OECD PISA **Economic, Social, and Cultural Status(ESCS) composite index.**

ESCS is a standardized latent index incorporating:

- parental education,
- parental occupational status,
- household possessions,
- cultural resources,
- educational resources in the home.

Its continuous scale allows high-resolution differentiation of structural socioeconomic conditions.

#### B.1.2. Outlier Removal Procedure

To prevent extreme ESCS values from distorting percentile ranks and the correction rule, outliers are removed using the standard:

$$1.5 \times IQR \qquad (B.1)$$

criterion applied to the full Korean PISA 2022 distribution.

- **Original sample**: 6,391 examinees
- **Identified outliers**: 14
- **Final analytic sample**: 6,377 examinees

This step stabilizes the SES distribution and prevents artificial skewing of $\mu$ or percentile boundaries.

**Note on the Tukey Rule.**

Outliers were removed using the standard Tukey rule (values outside $Q1 - 1.5 \cdot IQR$ or $Q3 + 1.5 \cdot IQR$). This criterion is distribution-free and robust to skewness, making it appropriate for educational assessment data where SES distributions often exhibit non-normality.

#### B.1.3. Percentile Normalization

To align SES with the correction rule and ensure interpretability across cohorts, we transform ESCS into a percentile rank (**Eq. (7)**):

$$S_i = PercentileRank(ESCS_i)$$

where,

- $Si \in [0,1]$
- $S_i$ is approximately uniform on [0,1] in large samples
- $S_i = 0$ denotes the lowest-SES student in the cohort
- $S_i = 1$ denotes the highest-SES student

This percentile normalization yields several analytical advantages:

1. **Comparability**: SES becomes comparable across cohorts and institutions.
2. **Transparency**: Quartile-based reporting becomes direct and interpretable.
3. **Distribution-aware correction**: Because a percentile-rank SES has $E[S_i] = 0.5$ (**Eq. (3)**), the correction rule naturally targets **the structurally disadvantaged bottom 50%.**

#### B.1.4. Population-Level vs. Applicant-Pool Percentiles

A critical design requirement for AMF is that SES percentiles must be computed relative to the **national population**, not the applicant pool.

**Why This Matters:**

In the PISA 2022 Korea dataset, the sample (N=6,377) is nationally representative by design, ensuring that the empirical mean $E[S_i] \approx 0.5$ (**cf. Eq. (3)**) However, in real admissions contexts, applicant pools are often **self-selected and skewed toward higher SES**.

For example:

- Institutions with affluent applicant pools: Applicant pool may be 60th-90th percentile nationally → pool mean $\mu \approx 0.75$
- Institutions with disadvantaged applicant pools: Applicant pool may be 20th-50th percentile nationally → pool mean $\mu \approx 0.35$

**Consequences of Applicant-Pool Normalization**:

If SES percentiles are computed within the applicant pool rather than nationally:

1. **Affluent applicant pool case ($\mu = 0.75$)**

- A student at the 60th national percentile becomes "bottom 10%" within the applicant pool
- Correction formula (**Eq. (4)**) gives them $S_i \approx 0.1$ (pool rank) → large correction

2. **Disadvantaged applicant pool case ($\mu = 0.35$)**

- A student at the 40th national percentile becomes "top 50%" within the applicant pool
- Correction formula gives them $S_i \approx 0.6$(pool rank) → negative correction (none)
- **Problem**: This student is disadvantaged nationally but receives no correction

**Strategic Manipulation Risk:**

If percentiles are pool-relative, institutions could manipulate outcomes by selectively recruiting high-SES applicants to raise the pool mean, thereby making moderate-SES students appear "disadvantaged" and inflating correction eligibility.

**Solution**:

AMF is intended to be anchored to **national administrative data** (e.g., census, tax records, education ministry databases) rather than computed within each institution's applicant pool.

This implies:

- $\mu = 0.5$ consistently across all institutions
- Correction targets genuine structural disadvantage
- No gaming via selective recruitment

- Cross-institutional comparability

**Implementation Mechanism**:

When a student applies, their SES index (income, parental education, region) is submitted to a national database API (e.g., Korea's National Tax Service, U.S. Department of Education). A feasible implementation would have an API return their **national SES percentile** ($S_i$ relative to the entire age cohort), not their rank within the university's applicant pool. This percentile is then used in the correction formula.

**PISA 2022 Korea as Validation**: The PISA sample approximates this ideal: it is nationally representative, so percentile ranks computed within the sample closely match national ranks. This is why $\mu \approx 0.5$ holds empirically and why the correction rule (**Eq.** (**4**)) functions as intended. Real implementations must preserve this property through external data integration.

### B.1.5. Quartile Assignment

Quartiles are defined with respect to the **entire analytic population**, not within the set of additional admits.

$$Q1: \ 0.00 \leq Si < 0.25$$

$$Q2: \ 0.25 \leq Si < 0.50$$

$$Q3: \ 0.50 \leq Si < 0.75$$

$$Q4: \ 0.75 \leq Si \leq 1.00$$

This implies:

- **consistent interpretation** of SES strata
- **no distortion** caused by computing quartiles inside small subsets (e.g., additional admits)
- **accurate analysis of SES targeting, especially when reporting Q1/Q2-only selection patterns under AMF**

### B.1.6. Connection to the Correction Rule

Since $S_i$ is percentile-normalized as **Eq. (3)**, the correction rule simplified from **Eq. (2)** to **Eq. (4).**

With the non-negativity constraint $C_i = \{\alpha \cdot (0.5 - S_i), 0\}$ , it follows that:

- only applicants in **Q1 and Q2** receive non-zero correction
- **Q3 and Q4 are structurally excluded**
- the empirical pattern "**100% of additional admits originate from the bottom 50%" is an immediate mathematical consequence of normalization, not a policy-imposed rule.** A full derivation and implications of this property appear in **Appendix C**.

### B.2. Empirical Calibration of the Policy Parameter $\alpha$

This appendix provides the empirical rationale for selecting the policy intensity parameter $\alpha \in \{5,10,15\}$ and documents the magnitude of corrections induced by these values within the PISA 2022 Korea dataset.

#### B.2.1. Estimating the SES-Achievement Gradient

We quantify the magnitude of structural disadvantage by regressing mathematics performance on the ESCS index. Let $R_i$ denote standardized math scores and let $ESCS_i$ denote raw socioeconomic status. (In the correction framework of Section 3, the same performance measure is denoted $M_i$)

$$R_i = \beta \cdot ESCS_i + \varepsilon_i \qquad (B.2)$$

For the 2022 Korea sample, when $\beta = 47.29, p < 0.001, R^2 = 0.136$.

Thus, a one-standard-deviation increase in ESCS ($\sigma_{ESCS} = 0.823$) corresponds to:

$$\Delta \approx 47.29 \times 0.823 = 38.90 points$$

This figure provides an empirical benchmark for the "environmental effect" of socioeconomic status.

#### Note on ESCS Standard Deviation

Although ESCS is standardized socioeconomic index (mean 0, SD 1) at the population level, the standard deviation can deviate from 1 when computed on a specific country's sample without applying PISA sampling weights. In this study, we use the raw Korean PISA sample (N=6,377), whose empirical ESCS SD is 0.823. Using the sample SD is methodologically consistent with the regression coefficient $\beta$ estimated on the same dataset.

### B.2.2. Design Principle for $\alpha$

To ensure that AMF remains conservative relative to documented SES effects, we calibrate $\alpha$ such that the *maximum* possible correction does not exceed a modest fraction of the SES-achievement gradient.

Given the empirical correction rule (**Eq. (4)**), the maximum correction occurs at $S_i = 0$:

$$C_{max} = 0.5 \cdot \alpha \tag{B.3}$$

The tested values therefore imply:

**Table B.1. Calibration of Policy Intensity Parameter $\alpha$**

| $\alpha$ | **Maximum correction** | **Fraction of SES effect** |
|---|---|---|
| **5** | 2.5 pts | 6% |
| **10** | 5.0 pts | 13% |
| **15** | 7.5 pts | 19% |

Even the highest setting ($\alpha$ = 15) corrects **less than one-fifth** of the empirically estimated SES impact on performance.

This calibration is designed such that AMF corrects for structural disadvantage without overcompensating or elevating low-performing students above competitive thresholds.

**Choice of $\alpha$ Range.**

The SES-achievement relationship in the Korean PISA sample shows that a one-standard-deviation increase in ESCS corresponds to a 38.90-point increase in mathematics scores ($\beta$ = 47.29, $\sigma_{ESCS}$ = 0.823). To avoid over-correction while still reflecting measurable socioeconomic constraints, $\alpha$ was chosen as a conservative fraction (6-19%) of this empirically estimated SES effect. This range implies that the AMF correction is anchored to observed data rather than normative assumptions, while keeping the adjustment modest relative to the underlying achievement gradients.

### B.2.3. Observed Correction Magnitudes in PISA 2022 Korea

The empirical correction magnitudes in the Korean sample ($N$ = 6,377) fall within narrow ranges:

**Table B.2. Observed Correction Magnitudes Among Conditional Admits (PISA 2022 Korea)**

| $\alpha$ | Min $C_i$ | Max $C_i$ | Mean $C_i$ |
| --- | --- | --- | --- |
| **5** | 0.99 | 2.32 | ~ 1.48 |
| **10** | 1.98 | 4.64 | ~ 2.99 |
| **15** | 2.97 | 6.95 | ~ 4.76 |

These magnitudes indicate that:

- AMF provides **small, distribution-aware adjustments**
- corrections are **unlikely to** elevate low-ability applicants
- only **near-threshold, structurally disadvantaged** students can cross the merit cutoff

#### B.2.4. Interpretation and Policy Implications

This calibration strategy yields three important properties:

1. **Proportionality**

   Corrections scale linearly with structural disadvantage but remain strictly bounded.

2. **Transparency**

   Institutions can choose $\alpha$ according to their desired level of opportunity expansion while maintaining merit anchoring.

3. **Predictable behavior**

   Because maximum corrections are tightly bounded, increases in $\alpha$ result in **modest, stable** changes in admit counts (e.g., 4 → 6 → 9 under $\alpha$ = 5/10/15).

These properties collectively imply that AMF functions as a **conservative yet meaningful** fairness mechanism rather than a redistributive tool.

## Appendix C. Formal Properties, Threshold Geometry, and Selection Logic

### C.1. Derivation of the Linear SES-Based Correction Rule

Let $M_i$ denote applicant *i*'s raw performance score and $S_i \in [0,1]$ denote the normalized SES index. AMF adjusts performance using a linear function that preserves relative SES differences while ensuring a transparent and auditable mapping from SES to correction magnitudes.

#### C.1.1. Correction Rule (Recap and Properties)

The linear correction rule and its percentile-normalized form are established in Section 3 (**Eqs. (1)–(4)**). Under percentile normalization, $S_i \in [0,1]$ with $E[S_i] = 0.5$ (**Eq. (3)**), and the correction reduces to $C_i = \alpha \cdot (0.5 - S_i)$ (**Eq. (4)**). This formulation satisfies three structural properties: (i) monotonicity—lower SES yields weakly higher correction; (ii) upper bound—corrections vanish at and above the median; and (iii) scale invariance—any affine transformation of SES is absorbed into the distributional center. The formal consequences of these properties are derived below.

#### C.1.2. Eligibility Boundary

An immediate consequence is that:

$$Ci > 0 \Leftrightarrow Si < 0.5$$

$$Ci = 0 \Leftrightarrow Si \geq 0.5$$

Thus, **approximately 50% of applicants are eligible for non-zero correction**, a property induced by percentile-normalization–not by policy choice. This explains the empirical observation that **100% of additional admits in the PISA analysis originate from the bottom 50% of SES.**

#### C.1.3. Corrected Performance and Threshold Geometry

Corrected performance is $M_i^* = M_i + C_i$ (**Eq. (5)**). Let T denote the raw-score threshold corresponding to the top-k admits. Because T is determined solely from raw scores, regular admits cannot be displaced.

Applicants qualify as conditional admits when $M_i^* \geq T \Leftrightarrow M_i + \alpha \cdot (0.5 - S_i) \geq T$ (**Eq. (4)-(6)**). Rearranging yields the minimum raw score required for a given SES:

$$M_i \geq T - \alpha \cdot (0.5 - S_i)$$

This establishes:

- **Linear threshold shift**: low-SES applicants require slightly lower raw scores, proportional to SES disadvantage
- **Binding constraint**: no applicant with $M_i < T - \alpha \cdot (0.5 - S_i)$ can be admitted
- **Merit preservation**: if $M_i$ is far below threshold, no correction can elevate the applicant above $T$

#### C.1.4. Bound on Maximum Correction

The maximum correction is $0.5 \cdot \alpha$ (**Appendix B.2.2**), yielding bounds of 2.5, 5.0, and 7.5 for $\alpha = 5, 10, 15$. Respectively, these bounds indicate that AMF operates within a narrow adjustment range, consistent with the empirical SES-achievement gradient observed in PISA.

#### C.1.5. Implication

The derivations above establish that:

- The correction rule is a **distribution-aware linear compensator.**
- Median SES ($S_i$ = 0.5) constitutes a **structural eligibility boundary**, not policy-imposed cutoff.
- Corrections cannot elevate low-performing applicants into competitive ranges.
- Thresholds remain fixed, implying **no displacement** of regular admits.

A full characterization of threshold geometry and sensitivity analysis appears in **Appendix D**.

### C.2. Merit Threshold Construction and Geometry

#### C.2.1. Definition of the Merit Threshold

For each cohort, the merit threshold $T$ is defined as the raw performance score of the:

$$k\text{–}th\ highest\ student$$

where,

- $k = [0.10 \times N_{app}]$ for top 10% admissions (Korean PISA example)
- $N_{app} = 6{,}377$ after outlier removal

- Thus $k$ = 638

Formally:

$$T = M_{(k)} \qquad (C.1)$$

where, $M_{(k)}$ denotes the $k$-th order statistic of the raw score distribution.

### C.2.2. Why AMF Does Not Modify the Threshold

AMF retains **the exact same merit threshold** used in existing admissions.

This design preserves:

- **non-displacement** of regular admits
- **merit anchoring**
- **compatibility** with existing institutional rules that tie admissions to quartile-based test scores

Mathematically, $M_i^* = M_i + C_i,\ but\ T = fixed\ from\ raw$ (**Eq. (5)**). Thus, AMF only asks a counterfactual question: "Would this applicant have exceeded the existing threshold if structural SES disadvantage had been equalized?"

### C.2.3. Threshold Geometry and Competitive Range

Define:

$$\Delta_i = T - M_i \qquad (C.2)$$

as the raw-score distance from the merit threshold.

Applicants eligible for AMF conditional admission must satisfy:

$$M_i^* = M_i + C_i \geq T \Leftrightarrow C_i \geq \Delta_i$$

this implies the structural condition:

$$\alpha \cdot (0.5 - S_i) \geq T - M_i$$

since **Eqs. (5)-(6), (C.2)**.

Only applicants who are both:

1. **close to the threshold** (small $\Delta_i$)

2. **structurally disadvantaged** (low $S_i$) can satisfy this inequality.

This formalizes one of AMF's key architectural features:

- High-SES applicants near the threshold cannot qualify.
- Low-SES applicants with very low raw scores cannot qualify.
- Only the **intersection zone (near-threshold × low-SES)** is eligible.

This is the mathematical reason PISA results show:

- 100% of additional admits originate from Q1-Q2.
- Additional admits always number small (4-9 out of 6,377).
- The model exhibits **selective, conservative expansion**.

**C.2.4. Stability Under Score-Distribution Changes**

Order-statistic thresholds such as $M_{(k)}$:

- scale smoothly with changes in overall score variance
- shift upward or downward with cohort competitiveness
- remain robust to small perturbations

Hence, even when robustness simulations inject SES noise or score variance (**Appendix D**), the binding condition $M_i^* \geq T$ (**Eq. (6)**) and the selection geometry remain stable.

This explains empirical findings:

- **Q3/Q4 representation remains minimal under noise perturbations** (Q3 ~1% in affected runs; Q4 = 0%)
- **threshold gaps remain positive for all additional admits**
- **correction magnitudes remain moderate and bounded**

**C.2.5. Connection to the Two-Phase Mechanism**

The threshold defined here feeds directly into the selection rule described in **Appendix C.2**:

- **Phase 1**: select raw-score top-$k$
- **Phase 2**: select all with $M_i^* \geq T$

The threshold plays a dual role:

- anchors merit standards
- acts as a fairness gate that only structurally disadvantaged, near-threshold applicants can pass

## Appendix D. Extended Results and Simulation Framework

### D.1. Simulation Protocol for PISA 2022 Korea

This appendix documents the full pipeline used to simulate AMF under the 2022 Korea PISA mathematics dataset. All steps correspond exactly to the empirical procedure summarized in **Section 3.4**.

#### D.1.1. Data Preparation

**(1) Raw dataset**

- Source: PISA 2022 Korea mathematics assessment
- Initial sample size: **6,391** students

**(2) SES outlier removal**

Outliers in ESCS are removed using the $1.5 \times IQR$ (**Eq. (B.1)**):

- Removed: **14** cases
- Final analytic sample: **6,377**

**(3) SES normalization**

SES is transformed into a percentile rank (**Eq. (7)**). This implies:

- an approximately uniform marginal distribution
- cross-cohort comparability
- an empirical mean close to 0.5 (up to rounding and sampling variation)

All subsequent corrections rely on this percentile-normalized SES.

#### D.1.2. Correction Computation

For each policy intensity $\alpha \in \{5,10,15\}$:

1. **Compute correction** (**Eq. (4)**, with non-negativity constraint $C_i = max\{C_i, 0\}$).
2. **Compute corrected score** (**Eq. (5)**).
3. **Record correction magnitude and SES distribution** for all eligible students $S_i < 0.5$.

#### D.1.3. Threshold Identification

The merit threshold $T$ is defined as the raw score of the top 10%. So, with **Eq. (C.1)**,

$$T = M_{(k)}, \qquad k = [0.10 \cdot 6377] = 638$$

Thresholds are computed **solely from raw scores,** ensuring:

- **no displacement** of regular admits
- direct comparability with existing admissions systems
- stability across $\alpha$ value

#### D.1.4. Conditional Admit Selection

Applicants qualify for conditional admission when $M_i^* \geq T$ (**Eq. (6)**).

For each $\alpha$:

- count number of conditional admits
- compute their SES quartile distribution
- compute threshold-gap values $M_i^* - T$
- record raw-score distances $\Delta_i = T - M_i$ (**Eq. (C.2)**)

This yields the key empirical results reported in the main text.

#### D.1.5. Robustness Procedures

We evaluate the stability of AMF under three perturbation classes.

(1) **SES Noise Robustness**

Inject noise $\epsilon_i \sim N(0, \sigma * 2)$ into ESCS values before percentile conversion and re-run the full pipeline:

- $\sigma$ = 0.05
- $\sigma$ = 0.10

(2) **Score Variance Perturbation**

Add noise to raw scores:

$$M_i' = M_i + \eta_i \tag{D.1}$$

with:

- $\eta_i \sim N(0, 5^2)$
- $\eta_i \sim N(0, 10^2)$

(3) **Threshold-Shift Analysis**

Adjust $T$ by ±1, ±2 points and measure sensitivity in additional admits.

Across robustness settings, the SES composition of conditional admits remains predominantly within **Q1-Q2**. Under SES noise perturbations, minor Q3 penetration (~1% mean share) occurs due to boundary misclassification, while Q4 remains at 0%. This supports the architectural stability of the mechanism.

### D.1.6. Long-Run Dynamics (DBN Integration)

AMF's short-run corrections feed into a **Dynamic Bayesian Network (DBN)** model that simulates multi-year mobility trajectories.

At each time step $t$:

- structural SES exposure influences latent ability
- corrected opportunity modifies future performance states
- transition probabilities update

Details of the DBN architecture, state definitions, and transition specifications appear in **Appendix E**.

## D.2. Extended Empirical Results for AMF (PISA 2022 Korea)

This appendix reports full empirical results underlying Section 4. All computations use the PISA 2022 Korea mathematics dataset ($N$ = 6,377), processed according to the protocol in **Appendix D.1**.

### D.2.1. Correction Distributions

Correction magnitudes **Eq. (4)** are strictly non-negative and bounded. **Table D.1** summarizes their empirical distributions among conditional admits (students who cross the threshold due to correction). Distributions are linear rescaling of one another, reflecting the structure of the correction function.

**Table D.1. Correction Distribution Among Conditional Admits**

| $\alpha$ | Min | Max | Mean | SD |
|---|---|---|---|---|
| **5** | 0.99 | 2.32 | ≈1.48 | ≈0.61 |
| **10** | 1.98 | 4.64 | 2.99 | 0.97 |
| **15** | 2.97 | 6.95 | 4.76 | 1.40 |

### D.2.2. Threshold-Gap Geometry

For each $\alpha$, additional admits satisfy with **Eq. (5)**:

$$M_i^* = M_i + C_i \geq \mathrm{T} = 666.62$$

**Table D.2** summarizes the corrected score gaps.

**Table D.2. Threshold Gap Distribution (Conditional Admits)**

| $\alpha$ | Min Gap | Max Gap | Mean Gap | SD |
|---|---|---|---|---|
| **5** | +0.23 | +1.50 | +0.73 | ~0.52 |
| **10** | +0.34 | +3.82 | +1.67 | ~1.16 |
| **15** | +0.16 | +6.14 | +2.48 | ~1.78 |

Threshold exceedance ranges remain small, indicating that AMF corrects only near-threshold underestimation.

### D.2.3. Raw-Score Distance Before Correction

The raw distance from the merit cutoff defined as $\Delta_i = \mathrm{T} - M_i$ (**Eq. (C.2)**). Additional admits originate from a narrow band of $\Delta_i$, as shown in **Table D.3.**

**Table D.3. Raw Score Distance ($\Delta_i$) Among Conditional Admits**

| $\alpha$ | **Min** $\Delta_i$ | **Max** $\Delta_i$ | **Mean** $\Delta_i$ |
|---|---|---|---|
| **5** | 0.23 | 1.50 | ~0.73 |
| **10** | 0.34 | 3.82 | ~1.67 |
| **15** | 0.16 | 6.14 | ~2.48 |

As $\alpha$ increases, the eligible raw-score radius expands moderately, but remains bounded.

### D.2.4. SES Distribution of Conditional Admits

SES composition is defined using cohort-level quartiles of the percentile-normalized SES index. **Table D.4** shows quartile shares. For all $\alpha$, additional admits originate exclusively from the bottom 50% of SES.

**Table D.4. SES Quartile Composition of Conditional Admits**

| $\alpha$ | **Q1** | **Q2** | **Q3** | **Q4** |
|---|---|---|---|---|
| 5 | 50% | 50% | 0% | 0% |
| 10 | 67% | 33% | 0% | 0% |
| 15 | 78% | 22% | 0% | 0% |

### D.2.5. Joint SES-Score Structure

To assess selection geometry, we compute the joint distribution of:

- percentile SES $S_i$
- raw score $M_i$
- corrected score $M_i^*$

Conditional admits cluster within a narrow, low-SES, near-threshold region:

- $S_i$ ranges:

– $\alpha = 5$: 0.04-0.30

– $\alpha = 10$: 0.04-0.30

– $\alpha = 15$: 0.04-0.30

- Raw scores:

– $\alpha$ = 5: 665.3-666.2

– $\alpha$ = 10: 664.1-666.2

– $\alpha$ = 15: 662.1-666.2

No high-SES or low-ability student appears near the corrected threshold, even at $\alpha = 15$.

**D.2.6. Effect of Increasing $\alpha$**

Increasing $\alpha$ expands conditional admits modestly:

- 4 → 6 → 9
- The set remains tightly concentrated in Q1-Q2
- Threshold-gap and $\Delta_i$ bands expand mildly but remain small

This behavior is consistent with the **conservative calibration** of $\alpha$ (**Appendix B.2**).

**D.2.7. Robustness Results**

Across robustness procedures (**Appendix D.3**):

1. **SES noise perturbation**

Quartile composition remains predominantly Q1-Q2, with Q3 appearing in approximately 15% of runs (mean share ~1%) due to boundary misclassification. Q4 remains at 0%. Corrections shift slightly, but overall targeting remains stable.

2. **Score variance perturbation**

Noise widens the raw-score band slightly, but threshold gaps remain positive and small.

3. **Threshold shift tests**

±1 − 2point adjustments cause predictable, monotonic changes in additional admits, without altering SES targeting.

Overall, AMF exhibits **stable behavior** across the tested perturbations.

**Summary**

The extended empirical results support three conclusions:

1. Selective identification of structurally disadvantaged, near-threshold applicants

2. Bounded, predictable corrections under all $\alpha$ settings

3. Stable functioning under noise and perturbation scenarios

These findings reinforce AMF's design goal of combining **merit preservation** with **structurally targeted opportunity expansion.**

### D.3. Robustness Checks: Perturbation Models and Stability Analysis

This appendix summarizes robustness tests evaluating whether AMF's behavior remains stable under changes to SES inputs, score variance, and threshold location.

All procedures follow the simulation pipeline in **Appendix D.1**.

#### D.3.1. SES Noise Perturbation

To examine sensitivity to measurement error in socioeconomic status, we perturb the ESCS index before percentile normalization:

$$ESCS_i' = ESCS_i + \epsilon_i,\ \epsilon_i \sim N(0, \sigma^2) \qquad (D.2)$$

with:

- $\sigma \in \{0.05, 0.10\}$

After perturbation, SES is re-normalized into percentiles via **Eq. (7)**, yielding $S_i' = PercentileRank(ESCS_i')$.

**Findings**:

Across all $\alpha$:

- Q1-Q2 composition of conditional admits remains dominant (~99% on average).

- Q3 appears in approximately 15% of simulation runs, with mean share ~1%. Q4 remains at 0%.

- This boundary leakage reflects symmetric misclassification: students near SES = 0.5 may shift across the Q2/Q3 boundary due to noise, causing some true-Q3 students to receive corrections and some true-Q2 students to miss them.

- Additional admits shift slightly in count (±1), but the SES profile remains stable overall.

- Threshold gaps remain positive (no borderline reversals).

**Reasons**:

Percentile normalization anchors rankings to the full distribution, so noise-induced shifts are typically small. However, students near the median (SES ≈ 0.5) may cross the Q2/Q3 boundary under measurement error. This explains why Q3 penetration is limited to ~1% on average: only boundary-region students are affected, and the effect is symmetric (some Q2 students also lose eligibility). Students far from the median—whether in Q1 or Q4—remain correctly classified with high probability.

### D.3.2. Raw-Score Variance Perturbation

We test AMF stability under score distribution changes by scaling the variance of the raw score distribution while preserving its mean:

$$M_i' = \left(M_i - \underline{M}\right) \times \sqrt{s} + \underline{M} \qquad (D.3)$$

with variance scale factors:

- $s \in \{0.8, 1.0, 1.2\}$ (20% decrease, baseline, 20% increase)

This simulates changes in test difficulty or applicant pool composition.

**Findings** (for $\alpha = 10$):

- Under reduced variance ($s = 0.8$), additional admits increase by 1 (from 6 to 7).
- Under increased variance ($s = 1.2$), the number of conditional admits remains stable at 6.
- The set of conditional admits remains drawn exclusively from Q1-Q2 across all scenarios.
- No high-SES student crosses the threshold in any perturbation scenario.

The mechanism's targeting precision persists even when the score distribution is perturbed.

**Reason**:

The correction envelope (≤ 7.5 points) remains small relative to the score variance; thus, noise does not elevate mid-performing Q3-Q4 students into the competitive band.

### D.3.3. Threshold-Shift Analysis

To assess sensitivity to institutional changes in merit standards, we test AMF under different threshold percentiles (top 5%, 10%, and 15% cutoffs), simulating varying levels of institutional selectivity.

**Findings**:

- Top 5%: 6 additional admits (threshold = 698.43)
- Top 10%: 6 additional admits (threshold = 666.62, baseline)
- Top 15%: 8 additional admits (threshold = 642.94)
- SES targeting remains strictly within Q1-Q2 across all threshold levels.
- No scenario introduces Q3-Q4 admits.

**Reason**:

The intersection of (low-SES × near-threshold) remains narrow; threshold changes only translate the binding boundary slightly without altering its geometry.

#### D.3.4. Stability of the Targeting Mechanism

Across all perturbations:

**(1) SES targeting remains robust**

Under score variance and threshold perturbations:

$$Q1 - Q2 \ \rightarrow \ 100\%,$$

$$Q3 - Q4 \ \rightarrow \ 0\%.$$

Under SES noise perturbations:

$$Q1 - Q2 \ \rightarrow \sim 99\%,$$

$$Q3 \ \rightarrow \sim 1\% \text{ (boundary misclassification)},$$

$$Q4 \ \rightarrow \ 0\%.$$

**(2) Correction magnitudes remain within designed bounds**

Maximum correction remains $\alpha/2$.

**(3) Threshold-gap signs remain positive**

No corrected score falls below the threshold ex post.

**(4) Additional admits remain few**

The model's conservative selection behavior is preserved.

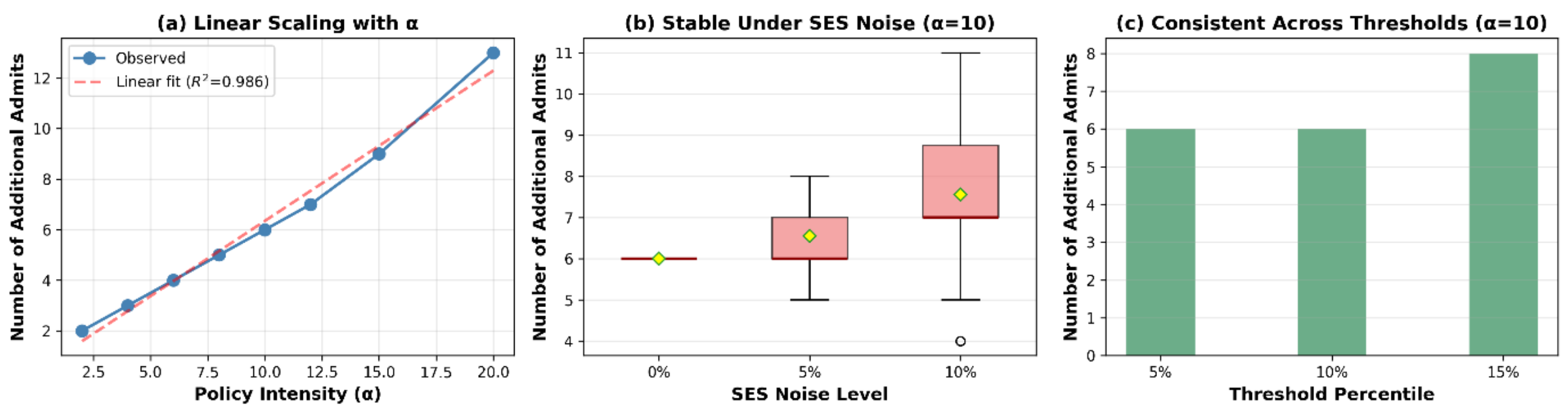

**Figure D.2. Robustness to Perturbations.** (a) Linear scaling of additional admits with $\boldsymbol{\alpha}(\boldsymbol{R^2} = \mathbf{0.986})$. (b) Stability under 5-10% SES noise ($\boldsymbol{\alpha} = \mathbf{10}$). (c) Consistent targeting across threshold percentiles (5%, 10%, 15%).

### D.3.5. Interpretation

The robustness tests support the structural stability of AMF's core behavior—**selective expansion, strict merit preservation, and robust SES targeting**.

This stability arises from three design features:

1. **Percentile SES normalization** Ensures the bottom 50% remains the primary correction-eligible region. Under measurement noise, boundary students (near SES = 0.5) may be misclassified, but this effect is symmetric and limited in magnitude (~1% Q3 penetration on average).

2. **Small correction envelope** Prevents low-ability applicants from approaching the threshold.

3. **Fixed merit threshold** Maintains consistency across perturbations.

Together, these properties imply that AMF functions as a conservative, predictable mechanism even under substantial perturbation to SES or performance inputs. The minor Q3 leakage under SES noise does not compromise the mechanism's core targeting logic but reflects an inherent trade-off in any system relying on imperfectly measured inputs.

### D.4. Population-Weighted Re-aggregation Analysis

The PISA microdata are sample-based and incorporate population weights to reflect national-level distributions. To examine how aggregate effect magnitudes change when sample results are re-weighted to reflect population structure, this study computes population-weighted estimates using the standard PISA student sampling weight (W_FSTUWT), and compares them to the unweighted sample results reported in **Section 4**.

Across all three values of $\alpha$, population weighting increases the absolute number of conditional admits but preserves the qualitative pattern: (i) all admits remain within the bottom half of the SES distribution, (ii) targeting precision remains approaches 100%, and (iii) the median SES of conditional admits remains well below the overall cohort median. The weighted distributions therefore amplify the magnitude of AMF's impact without altering the underlying selectivity structure.

**Fig. D.1** illustrates the sample versus population-weighted comparisons. While the sample counts (e.g., 9 admits at $\alpha$ = 15) are small due to the limited microdata, the weighted estimates correspond to realistic national-level projections (approximately 760 admits). These results suggest that although effect sizes appear modest in the raw sample, their policy-relevant magnitude is substantially larger once population structure is restored.

Taken together, the weighted analysis supports the stability of the main findings: AMF's correction rule remains selective, merit-preserving, and structurally targeted even when scaled to population-level estimates.

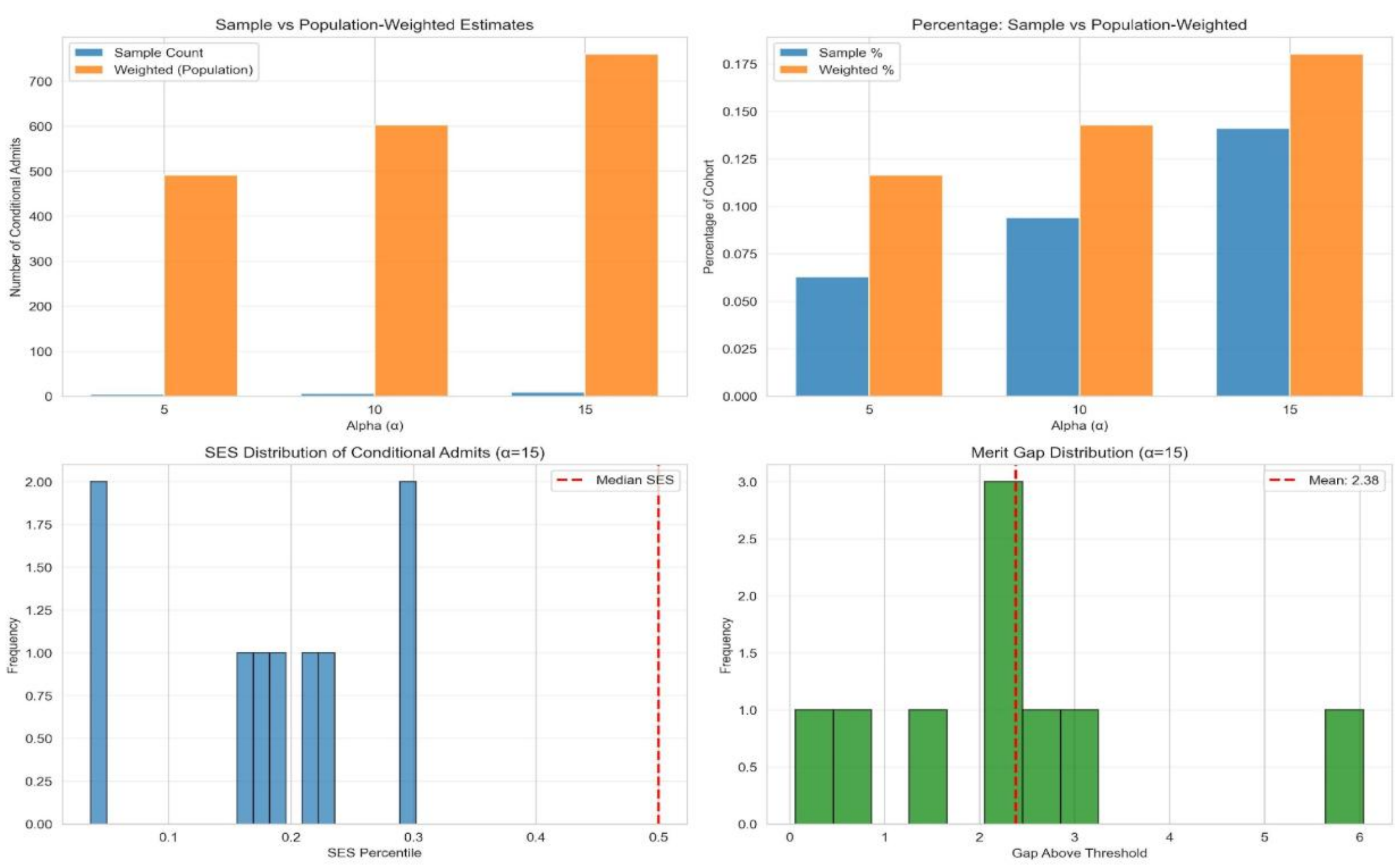

**Figure D.1. Comparison of sample counts and population-weighted estimates for conditional admits, along with SES and merit gap distributions.**

## Appendix E. Illustrative Dynamic Bayesian Network Framework for Long Run Mobility Simulation

This appendix outlines the dynamic Bayesian network (DBN) used to qualitatively explore the potential long-run implications of AMF. The DBN is not part of the core contribution but provides a conceptual simulation framework for studying how small, targeted corrections may propagate through multi-year educational trajectories. These latent states are used for simulation purposes only and are not statistically inferred from data.

### E.1. Model Overview

The DBN represents educational progression as a sequence of latent ability states $X_1, X_2, \cdots, X_T$. Where each $X_t$, represents an ordered latent academic ability tier at time t, used for simulation purposes.

Each student is associated with:

- **structural SES exposure** $S_i$ (exogenous)
- **opportunity correction** $C_i$ (nonzero only once, at t=1 under AMF)
- **performance proxy** $R_t$ (e.g., standardized test score indices, used illustratively)

The DBN illustrates how these quantities influence transitions from $X_t$ to $X_{t+1}$

### E.2. State Space

Let $X_t \in \{1,2,3,4\}$ denote ordered latent ability tiers.

These tiers correspond approximately to relative levels of academic readiness (e.g., coarse percentile groupings) but are not directly observable.

The simulation links latent states to stylized performance proxies. For example, a simple Gaussian mapping may be used for illustration:

$$R_t \sim N\left(\mu_{X_t}, \sigma^2_{X_t}\right) \qquad (E.1)$$

Parameter magnitudes may be loosely informed by cross-sectional distributions (e.g. PISA, PIRLS, TIMSS), without performing formal calibrations.

### E.3. Transition Model

Transitions follow a first-order Markov structure:

$$P(X_{t+1}|X_t, S_i, C_i) \tag{E.2}$$

We parameterize the transition kernel using three components:

(1) **Baseline mobility matrix** A row-stochastic matrix $M$ reflecting average year-to-year mobility:

$$M_{ab} = P(X_{t+1} = b|X_t = \alpha) \tag{E.3}$$

(2) **SES exposure term** SES percentile $S_i$ nudges transitions:

- upward mobility probability decreases linearly in $S_i$
- downward mobility probability increases accordingly

One stylized specification is:

$$M_{ab}^{(SES)} = M_{ab} + \sigma \cdot (S_i - 0.5) \cdot \gamma_{SES} \tag{E.4}$$

with small $|\gamma_{SES}|$.

This representation is illustrative and captures the directional effect of SES on mobility rather than an estimated structural relationship.

(3) **AMF correction term** The correction $C_i$ acts as an initial condition shift at $t = 1$:

- it increases the probability of entering higher latent states
- the effect decays over time (structural, not permanent)

One implementation:

$$M_{ab}^{(AMF)} = M_{ab}^{(SES)} + f(C_i) \cdot 1(t = 1, a < b) \tag{E.5}$$

where, $f(\cdot)$ is a bounded, increasing function (e.g., logistic or linear with saturation).

The combined transition kernel is normalized to maintain stochasticity.

**E.4. Mobility Matrix Calibration**

The transition matrices $M_{admit}$ and $M_{not}$ are stylized structures designed to reflect widely documented patterns in intergenerational mobility research: (a) higher persistence at top and bottom SES quartiles, and (b) modest upward mobility for middle groups.

The probabilities were constructed to satisfy four constraints:

1. Row-stochasticity (each row sums to 1)

2. Monotonic patterns in upward mobility as SES increases

3. Lower downward mobility for top-SES groups

4. Qualitative consistency with reported mobility trends in Korea's reported mobility trends [1]

The initial distribution $v_0$ = [0.35,0.30,0.20,0.15] reflects a stylized, moderately bottom-heavy SES structure inspired by recent Korean demographic patterns. This distribution serves as a baseline state for long-run dynamic analysis; it is not used for statistical inference.

These matrices are constructed to reflect qualitative patterns consistently reported in the intergenerational mobility literature: higher persistence at both tails of the SES distribution, lower downward mobility among top-SES groups, and moderate upward mobility for middle-ranked groups. These stylized features appear across multiple empirical contexts, including U.S. administrative mobility studies [22], OECD's cross-national mobility analyses [23], and recent Korean evidence [1].

### E.5. Simulation Dynamics

For each student:

#### Step 1 – Initialize $X_1$ from a stylized baseline distribution

Assign an initial latent state $X_1$ based on stylized initialization rule (not inferred from data).

#### Step 2 – Apply AMF at t=1

Compute $C_i$ and incorporate into the transition kernel only for the first step.

#### Step 3 – Propagate forward

Simulate $X_2, X_3, \cdots, X_T$ under the modified kernel.

**Step 4 – Generate long-run outcomes**

Common descriptive summaries include:

· long-run mean latent ability

· probability of reaching top states

· mobility indices (upward transitions vs. downward)

· inequality metrics across SES strata

### E.6. Long-run Convergence Properties

To examine the intergenerational impact of AMF, we simulate 30 generations under baseline and AMF scenarios. **Fig. E.1** illustrates the evolution of SES distribution over time. Under baseline policy (no intervention), the Q1 share converges to 39.5%. With AMF ($\alpha$ = 10), Q1 share converges to 39.5%, representing a 0.05 percentage point improvement (0.1% relative reduction). While the immediate long-run effect is modest, this is indicative of AMF's stability in the simulated setting: the mechanism expands opportunity without evidence of disrupting systemic equilibria or generating cascading distortions. The conservative design of AMF is intended to be compatible with existing institutional structures while maintaining potential for cumulative effects through sustained implementation across multiple cohorts.

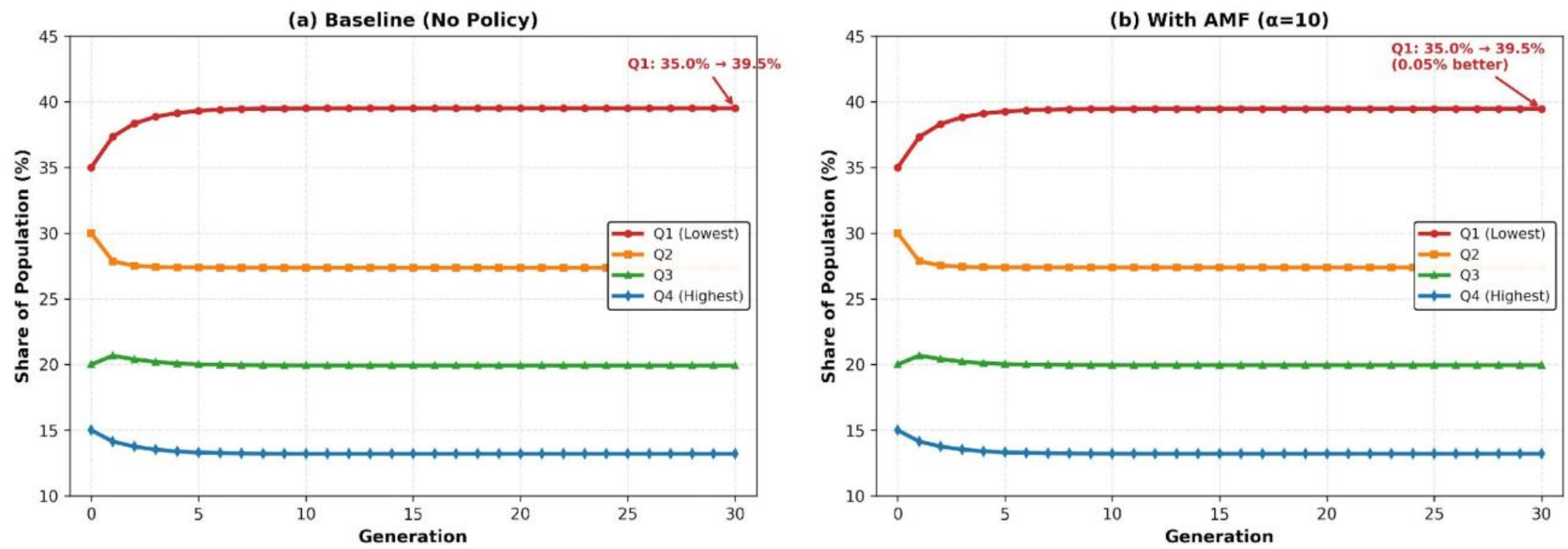

**Figure E.1. DBN Long-term Trajectories**

### E.7. Interpretation of DBN Outputs

Across implementations tested in the AMF engine:

- AMF induces a **small but observable upward shift** for low-SES students in the first 1-2 periods.
- Effects **decay naturally**, mitigating the risk of runaway amplification.
- High-SES students exhibit **minimal change**, preserving merit anchoring in the simulation.
- Cohort-level inequality metrics (e.g., top-state reach probability) shift modestly.

These qualitative patterns are broadly consistent with empirical research on mobility elasticity and with Chetty-style models of early-stage opportunity shocks.

### E.8. Connection to Main Text

The DBN does **not** affect any baseline simulation results.

It is included to show that AMF:

- is compatible with dynamic modeling,
- does not exhibit instability when propagated forward in the simulation, and
- yields long-run patterns qualitatively aligned with the literature on the scarcity of "hidden excellence" [17].

Full implementation details and simulation code are available in the public code repository (see **Code and Data Availability section**).

## Appendix F. Policy Implementation and Decision Modules

### Scope and Non-Interference Principle

This appendix specifies how the AMF can be implemented within existing admissions systems without altering its internal decision spine. While **Section 5** defines the invariant sequence of decision-making stages, **Appendix F** describes institutional conditions under which those decisions are operationalized.

The components discussed here function strictly as **implementation modules**. They do not introduce additional decision logic, modify applicant rankings, or alter the role of $\alpha$ as the sole normative policy lever. Instead, they clarify where AMF is embedded within existing admissions architectures, how fiscal and capacity constraints are accommodated, and how exceptional procedures may be activated without interfering with the decision spine.

Within this scope, the appendix accommodates distinct implementation conditions arising from fiscal and capacity constraints. These conditions give rise to different deployment configurations, which are formalized in the subsequent sections without introducing additional decision logic or normative parameters.

This separation ensures analytical transparency and auditability while allowing AMF to be adopted under heterogeneous legal, fiscal and institutional conditions.

### F.1. Entry Points and Integration Model

This section focuses on institutional integration and compatibility, rather than on the internal construction of merit or equity correction, which are specified in Section 5. It clarifies how the AMF interfaces with existing admission systems, and what may vary without changing **Section 5**.

AMF is designed as a **unified allocation mechanism**, not as an additional admissions track. Accordingly, its integration is based on **consolidation rather than coexistence.**

AMF is not intended to integrate directly with existing holistic admissions models that rely heavily on qualitative or portfolio-based evaluation (e.g., self-reported activities, essays, recommendation letters). Such components introduce substantial variability, are highly sensitive to differential access to information and private tutoring, and limit the transparency needed for merit-preserving corrections.

Instead, AMF is compatible with **future multi-metric systems** in which the evaluation structure is redesigned to be more measurable, diversified, and education-driven rather than portfolio-driven, without prescribing how institutions must construct such systems.

Examples of such evaluation structures include:

- standardized indicators of academic preparation (e.g., CSAT or equivalent),

- creativity or problem-solving assessments administered under standardized conditions,

- longitudinal measures of school-based engagement or persistence,

- domain-general basic competency tests aligned with intended majors (kept at accessible difficulty levels),

- optional bonus components for applicants with exceptional domain-specific talent (maintaining a separate talent track).

Such a framework reduces reliance on opaque, preparation-intensive components, encourages students to prepare along a unified track, and limits the influence of private tutoring markets. Within this redesigned evaluation space, AMF operates naturally on the composite merit index: providing transparent and continuous adjustments while preserving the integrity of the underlying metrics.

#### F.1.1. Unified-Pool Integration (Quota Consolidation Model)

In many admissions systems, equity objectives are implemented through multiple categorical quotas, such as region-based, income-based, or background-based admission tracks. These categories typically operate as discrete eligibility gates with separate seat allocations.

Under AMF, such categorical structures are **not preserved as allocation mechanisms**. Instead, the informational content underlying these quotas is absorbed into the disadvantage measurement process. Variables that previously determined categorical eligibility are treated as inputs to the disadvantage indicator set and incorporated through the aggregation stage of the decision spine.

As a result, all applicants—regardless of background—are evaluated within a **single unified pool**, ranked by corrected scores derived from a common decision rule. No applicant is assigned to a separate track, and no seats are pre-allocated to specific categories.

This consolidation transforms equity policy from a system of discrete classifications into a **continuous correction mechanism**, reducing fragmentation, administrative complexity, and categorical labeling effects.

#### F.1.2. Implications for Legacy Quotas

The transition to AMF does not require denying the policy rationales that motivated existing quotas. Rather, it re-encodes those rationales in a form compatible with rule-based, auditable decision making.

In AMF, region, income, or background considerations influence outcomes only through their contribution to disadvantage signals and the calibrated correction function. They no longer operate as independent admission rules or seat guarantees.

This distinction is central: **quotas are treated as data, not as institutional constraints.**

#### F.1.3. Relation to the Decision Spine

The integration model described here does not modify the internal structure of the AMF decision spine. Indicator definition, aggregation, equity calibration, execution, and closure proceed identically regardless of the legacy system from which inputs are drawn.

Institutional integration determines *where* AMF replaces existing mechanisms, not *how* AMF operates internally. This separation preserves both conceptual clarity and implementation flexibility.

### F.2. Capacity as a Feasibility Condition (Not a Decision Lever)

Capacity constraints specify whether a given equity calibration can be implemented in practice; they do not constitute an independent decision rule. Within AMF, capacity does not operate as a normative lever alongside $\alpha$, nor does it modify the internal decision spine. Instead, capacity functions as a feasibility condition that bounds which values of $\alpha$ can be realistically adopted by an institution.

Formally, $\alpha$ determines the sensitivity of socioeconomic correction within the decision rule, while capacity determines whether the resulting level of adjustment can be accommodated given institutional, fiscal, and instructional constraints. Capacity therefore affects *which $\alpha$ values are admissible,* not *how applicants are ranked* once $\alpha$ is chosen.

This distinction is critical. Treating capacity as an execution-stage constraint or as a post hoc adjustment mechanism would reintroduce discretionary intervention after decisions are made. AMF instead requires that feasibility considerations be incorporated ex ante, at the stage of $\alpha$ selection, ensuring that once execution begins, outcomes follow mechanically from the fixed decision rule.

In operational terms, institutions evaluate candidate values of $\alpha$ by considering their local applicant distributions, expected ranges of additional admits, per-student costs, and instructional capacity. These assessments are institution-specific and need not be standardized across universities. What is standardized is the principle that capacity considerations enter only at the level of $\alpha$ selection and never alter the decision logic itself.

Accordingly, capacity does not substitute for merit-based admissions, nor does it operate as a quota. If an institution anticipates that a given value of $\alpha$ would exceed feasible intake, the appropriate response is to select a lower $\alpha$ ex ante, rather than to apply ad hoc adjustments or reallocate admissions ex post. This preserves both merit anchoring and procedural integrity.

By locating capacity strictly as a feasibility condition on $\alpha$, AMF separates normative choice from operational constraint. The decision spine remains invariant, while institutions retain flexibility to adopt equity calibration at a scale consistent with their resources and mission.

### F.3. Budget-Linked $\alpha$ Selection (Support Functions)

Public authorities do not select $\alpha$ directly. Instead, their role is to define the fiscal conditions under which different values of $\alpha$ can be implemented. Within AMF, government intervention is therefore expressed through **budget-linked support functions**, rather than through direct constraints on the decision rule.

Specifically, central authorities specify the level and structure of financial support associated with alternative ranges of $\alpha$, including per-student subsidies, marginal funding for additional admits, or total expenditure ceilings. These commitments define the feasible implementation space within which institutions may choose $\alpha$, without prescribing a single value.

This approach reflects two structural realities. First, the realized impact of $\alpha$ depends on local applicant distributions and competitive density, which vary substantially across institutions and programs. Second, attempting to regulate $\alpha$ directly would require central authorities to anticipate institution-specific outcomes, reintroducing informational asymmetries and discretionary adjustment. Budget-linked support functions avoid this problem by shifting feasibility assessment to the institutional level.

Under this arrangement, institutions evaluate candidate values of $\alpha$ by combining centrally announced support conditions with local information on applicant distribution, instructional capacity, and per-student costs. The selection of $\alpha$ thus becomes a bounded institutional decision: universities are free to choose $\alpha$ within the support regime but must internalize the fiscal and operational consequences of that choice.

Importantly, this structure preserves the role of $\alpha$ as the sole normative policy lever within the decision spine. Public authorities influence the scale of equity calibration indirectly, by shaping the fiscal environment in which $\alpha$ is selected, rather than by modifying the decision logic itself. Once $\alpha$ is chosen, admissions outcomes follow mechanically from the fixed decision rule.

By separating fiscal commitment from decision execution, AMF enables transparent responsibility allocation. Governments are accountable for the support regimes they define; institutions are accountable for the $\alpha$ values they select within those regimes, and the resulting outcomes can be evaluated ex post against stated equity and budgetary objectives.

### F.4. Opt-In Boundaries and Information Disclosure

This subsection clarifies how opt-in mechanisms operate in relation to the decision spine and other implementation modules. Opt-in does not constitute a decision rule, nor does it modify ranking, calibration, or execution logic. Instead, opt-in defines information boundaries under which the decision spine is applied.

Two distinct opt-in decisions are recognized within AMF.

**Pre-spine opt-in (eligibility for correction).**

Prior to entering the decision spine, applicants may opt in to the disclosure of socioeconomic information used for disadvantage measurement. Exercising this opt-in determines eligibility for socioeconomic correction but does not guarantee any adjustment or admission outcome. Applicants who do not opt in are evaluated solely on merit scores, using the same ranking and execution procedures.

Once the application cycle begins, the opt-in choice is binding for that cycle and cannot be revised ex post. This ensures that eligibility conditions are fixed before any aggregation, calibration, or execution occurs.

**Post-spine opt-in (linkage to support mechanisms).**

After admission outcomes are finalized, admitted students may separately opt in to the use of relevant information for post-admission support mechanisms, such as financial aid, academic assistance, or mentoring programs. This opt-in affects only the allocation of support resources and has no bearing on admission decisions.

Importantly, the two opt-in stages are analytically and temporally separated. Pre-spine opt-in affects eligibility for correction within the decision spine, while post-spine opt-in governs downstream support allocation. Neither introduces individualized discretion, and neither permits retrospective modification of admission outcomes.

By locating opt-in boundaries outside the internal logic of the decision spine, AMF preserves procedural integrity while respecting individual autonomy and data governance constraints.

### F.4.1. Information-Revealing Effects and Strategic Considerations

The opt-in structure of AMF creates potential information-revealing dynamics that merit attention.

**Signaling Effects**

Applicants who opt in to SES disclosure signal their disadvantaged status to the institution. While this information is used only for correction purposes, institutions must ensure that opt-in status does not influence other admissions dimensions or create stigma effects.

**Strategic Non-Disclosure**

Higher-SES applicants have no incentive to disclose, as disclosure yields zero or negative corrections. This self-selection is intentional and reinforces targeting precision. However, moderate-SES applicants near the median may face strategic uncertainty: small corrections may not justify the perceived costs of disclosure.

**Institutional Responses**

Institutions may adapt to AMF by adjusting other admissions criteria or support allocations. To prevent such spillovers from undermining AMF's intent, the framework specifies that correction operates within a fixed decision spine and does not interact with other evaluation dimensions.

**Gaming and Verification**

Because SES is measured through administrative data rather than self-report, direct gaming is constrained. However, strategic behaviors such as income timing or asset restructuring remain theoretically possible. The verification requirements in **Section 5.2** and audit mechanisms in **Section 5.5** are designed to mitigate these risks.

These considerations do not alter the core mechanism but inform implementation choices regarding data governance, communication strategies, and monitoring protocols.

## F.5. Emergency and Exceptional Modules (Optional)

### F.5.1. Rationale and Scope

The decision spine defined in **Section 5** is designed to operate under normal institutional conditions, where disadvantage can be represented using persistent and administratively verifiable indicators. However, rare but severe system-wide disruptions—such as pandemics, large-scale examination cancellations, or legally recognized national emergencies—may temporarily invalidate the informational content of merit signals for specific cohorts.

Emergency modules are introduced to address such situations without altering the invariant structure of the decision spine. Their role is not to redefine merit or equity, but to provide a narrowly scoped, time-bounded adjustment mechanism when standard inputs fail to reflect opportunity in a meaningful way.

Crucially, emergency modules are exceptional by design. They are not intended to generalize beyond clearly defined events, nor to substitute for long-term socioeconomic correction.

### F.5.2. Design Principles

Emergency modules are governed by four design principles:

1. **Parallelism**

   Emergency-related adjustments enter the system as parallel administrative signals, not as modifications to the core SES aggregation or correction function. This preserves the integrity of the primary disadvantage signal.

2. **Temporality**

All emergency modules must be explicitly time-bounded. Each module applies only to cohorts directly affected by the identified disruption and expires automatically after the defined period.

3. **Verifiability**

Eligibility for emergency adjustment must rely on administratively verifiable criteria (e.g., cohort membership, exam-year classification, legally recognized disaster status), rather than individualized claims.

4. **Non-precedential Structure**

Emergency modules do not establish permanent categories of disadvantage. Their activation does not create standing entitlements beyond the specified execution cycle.

These constraints ensure that emergency responses do not erode the non-discretionary character of the decision spine.

### F.5.3. Implementation as a Parallel Signal

Formally, emergency modules operate as additive or multiplicative adjustments applied alongside the standard correction function, rather than replacing it. Let $C_i$ denote the standard AMF-corrected score for applicant $i$.

An emergency-adjusted score may take the form:

$$C_i^E = C_i + \beta \cdot E_i \tag{F.1}$$

where $E_i$ is a binary or bounded administrative indicator denoting eligibility under the emergency module, and $\beta$ is a fixed, pre-announced adjustment parameter.

When finer cohort differentiation is required—for example, distinguishing between multiple affected academic years—$E_i$ may be decomposed into multiple parallel indicators, each subject to the same transparency and sunset constraints.

Importantly, the emergency parameter $\beta$ is institutionally distinct from the equity calibration parameter $\alpha$. While $\alpha$ reflects structural equity objectives, $\beta$ reflects temporary informational degradation in merit measurement.

### F.5.4. Illustrative Scenarios

Emergency modules may be justified in limited scenarios such as:

- Cohorts whose terminal secondary education years coincided with prolonged school closures or exam cancellations.
- Applicants assessed under fundamentally altered evaluation regimes due to system-wide disruptions.
- Legally designated national emergencies with documented, cohort-specific educational impact.

By contrast, idiosyncratic personal hardship or localized incidents are not appropriate triggers for emergency modules and should be addressed, if at all, through existing socioeconomic indicators or post-admission support mechanisms.

### F.5.5. Relation to the Decision Spine

Emergency modules do not alter the ordering, sequencing, or irreversibility properties of the decision spine. They are instantiated only after indicator definition and aggregation, and prior to batch execution, operating within the same governance and audit boundaries as standard corrections.

Their inclusion therefore reinforces, rather than weakens, the core design objectives of AMF: to separate structural equity correction from ad hoc discretionary intervention.

### F.5.6. Sunset and Review

Every emergency module must include an explicit sunset clause and a post-execution review requirement. Continuation beyond the initial execution cycle requires renewed justification based on updated evidence of cohort-level distortion in merit signals.

Absent such justification, the module is automatically deactivated.

## F.6. Decision Highway: Integrated Admission-Support Pipeline (Illustrative Korean Context)

This subsection provides an illustrative application of the AMF implementation modules in a Korean institutional context. Its purpose is not to introduce additional decision logic, but to demonstrate how admission and post-admission support may be operationally linked through administrative automation while preserving the non-interference principle defined in **Appendix F**.

In many higher education systems, admissions and student support operate as institutionally and temporally separate processes. Admissions decisions are finalized first, while financial aid and academic support are allocated ex post through independent administrative channels. This separation creates two inefficiencies: (i) structurally disadvantaged high-potential applicants may fail to cross the admission threshold despite later support availability, and (ii) post-admission support mechanisms rely on redundant data collection and discretionary assessment.

The Decision Highway refers to an automated administrative pipeline that links AMF-based selection to post-admission support, without altering the internal decision spine or admission outcomes.

### F.6.1. Pre-Spine Opt-In and Eligibility Boundary

Prior to entering the decision spine, applicants are presented with a single pre-spine opt-in decision indicating whether they consent to the administrative verification of socioeconomic information for the purpose of AMF-based correction.

This opt-in does not guarantee correction or admission; it only defines eligibility for disadvantage-based adjustment if the applicant's raw merit score does not meet the cutoff.

Applicants who do not opt in are evaluated solely on merit scores under the same ranking and execution rules.

### F.6.2. Administrative Data Integration

In the Korean context, verified socioeconomic data already exist at the national level through the Korea Student Aid Foundation (KOSAF), which maintains income and asset percentile classifications based on administratively verified records. These data are conceptually aligned with the data governance requirements of AMF and may serve as validated substitutes for empirical proxies used in simulation studies.

Rather than constructing a new verification infrastructure, universities may interface with existing KOSAF data through batch-based administrative queries, subject to applicant consent and privacy safeguards.

### F.6.3. Batch Execution and Audit Closure

Once the decision spine proceeds to execution, all AMF-based corrections are applied through batch processing. No individualized exceptions or post hoc adjustments are permitted.

After execution, an independent expert oversight and audit body performs cross-validation between anonymized input records and finalized admission outcomes. Personally identifiable information (e.g., name, age, gender) is excluded; verification is conducted using anonymized identifiers and mapped administrative attributes.

Only upon confirmation of consistency is the system formally closed, and results are released. This audit-and-closure step reinforces irreversibility and protects against strategic manipulation.

#### F.6.4. Post-Admission Opt-In and Support Linkage

Following admission, a separate post-spine opt-in allows admitted students to consent to the use of verified socioeconomic data for post-admission support mechanisms, such as financial aid or academic assistance. This opt-in is analytically and temporally distinct from the admission-stage consent and does not affect admission outcomes.

Where institutional capacity permits, financial support may be implemented through net-pricing mechanisms (e.g., tuition reductions reflected directly in billing statements), rather than visible ex post transfers. This design reduces stigma, simplifies administration, and preserves horizontal equity by applying support eligibility to all low-SES students, regardless of whether they were admitted via AMF correction or raw merit.

#### F.6.5. Non-Interference with the Decision Spine

Throughout this integrated pipeline, the AMF decision spine remains invariant. Admission ranking, correction calibration, execution, and closure are unaffected by downstream support allocation.

The Decision Highway therefore functions as an administrative accelerator rather than a decision modifier: it links selection and support without reintroducing discretion, categorical labeling, or post hoc intervention.